\documentclass[journal]{IEEEtran}
\usepackage{amsmath,amssymb}
\usepackage{booktabs}
\usepackage{graphicx}
\graphicspath{{./images/}}
\usepackage{cite}

\usepackage{xcolor}

\usepackage{tikz}
\usetikzlibrary{arrows.meta}

\definecolor{cWheels}{HTML}{666666}
\definecolor{cRigid}{HTML}{00509B}
\definecolor{cElastic}{HTML}{E77B29}
\definecolor{cGRU}{HTML}{C8D317}
\definecolor{cGRUink}{HTML}{8A9200}
\definecolor{cInk}{HTML}{27313F}
\definecolor{cFill}{HTML}{F3F5F9}

\usepackage[per-mode=symbol]{siunitx}

\usepackage{microtype}
\usepackage{xspace}

\usepackage{transparent}

\newcommand{\mm}[1]{\boldsymbol{#1}}
\newcommand{\R}{\mathbb{R}}
\newcommand{\transpose}{^{\top}}

\newcommand{\tyes}{\checkmark}
\newcommand{\tpart}{(\checkmark)}
\newcommand{\tno}{--}
\newcommand{\rot}[1]{\rotatebox{90}{#1}}

\newcommand{\VehA}{Vehicle~A\xspace}
\newcommand{\VehB}{Vehicle~B\xspace}

\begin{document}

\title{A Hybrid UKF with Explicit Drivetrain Modeling\\
for Low-Velocity State Estimation of Electric Vehicles}

\author{Hendrik Sch\"afke,
        Daniel O.~M.~Weber,
		Thomas Seel,~\IEEEmembership{Member,~IEEE,}
		Askar Vagapov,
		Jannes Iatropoulos,
        Christoph Schweers,
        Ahmed Trabelsi,
        Roman Henze,
        and~Simon F.~G.~Ehlers%
\thanks{This work has been submitted to the IEEE for possible publication. Copyright may be transferred without notice, after which this version may no longer be accessible.}%
\thanks{This work was supported by IAV GmbH and the Institute of Automotive Engineering, Technische Universit\"at Braunschweig. \textit{(Corresponding author: Hendrik Sch\"afke.)}}%
\thanks{H.~Sch\"afke, D.~O.~M.~Weber, T.~Seel, and S.~F.~G.~Ehlers are with the Institute of Mechatronic Systems, Leibniz University Hannover, 30823 Garbsen, Germany. (e-mail: schaefke@imes.uni-hannover.de.)}%
\thanks{T.~Seel is also with the L3S Research Center, 30167 Hannover, Germany.}%
\thanks{J.~Iatropoulos and R.~Henze are with the Institute of Automotive Engineering, TU Braunschweig, 38106 Braunschweig, Germany.}%
\thanks{A.~Vagapov and A.~Trabelsi are with IAV GmbH, 38518 Gifhorn, Germany. C.~Schweers was affiliated with IAV GmbH, 38518 Gifhorn, Germany, when this work was performed.}%
}

\markboth{Preprint submitted to IEEE Transactions on Intelligent Vehicles}%
{Sch\"afke \MakeLowercase{\textit{et al.}}: A Hybrid UKF with Explicit Drivetrain Modeling}

\maketitle

\begin{abstract}
Longitudinal velocity is a core input for vehicle control, driver-assistance, and automated driving functions.
Production wheel-speed sensors deliver only a few pulses per revolution, so they lose resolution below 1\,km/h, the regime in which automated parking and traffic-jam assistance operate.
The motor speed from the electric machine's resolver stays finely resolved down to standstill but reaches the wheels only through a compliant drivetrain.
We therefore estimate the longitudinal velocity with an unscented Kalman filter (UKF) on a double-track model adapted to low velocities and incorporate the motor speed along two routes, one purely physical and one hybrid.
The first augments the model with an explicit torsional spring-damper drivetrain, making the motor speed a state of the estimator.
The second feeds the output of a neural network-based virtual wheel-speed sensor into the UKF as a pseudo-measurement.
Both routes are compared against two baselines, a wheel-speed-only configuration and one that adds the motor-speed sensor but treats the driveline as rigid.
Because the model is numerically stiff near standstill, the comparison rests on a systematic study of explicit and semi-implicit discretization schemes.
All four configurations are evaluated against an inertial reference in real-world test drives with two distinct series-production electric vehicles, anonymized as \VehA{} and \VehB{} for confidentiality.
Below 1\,km/h, both proposed estimators lower the velocity root-mean-square error by at least 54\,\% and 45\,\% against both baselines on \VehA{}.
Carried over zero-shot to \VehB{} without re-tuning on target-vehicle data, they lower it by at least 41\,\% and 51\,\%.
Both estimators therefore carry across vehicles without a new identification campaign.
\end{abstract}

\begin{IEEEkeywords}
Automated parking, Kalman filters, vehicle state estimation, virtual sensors.
\end{IEEEkeywords}

\section{Introduction}
\label{sec:introduction}

\begin{figure}[t]
\centering
{\footnotesize\def\svgwidth{\columnwidth}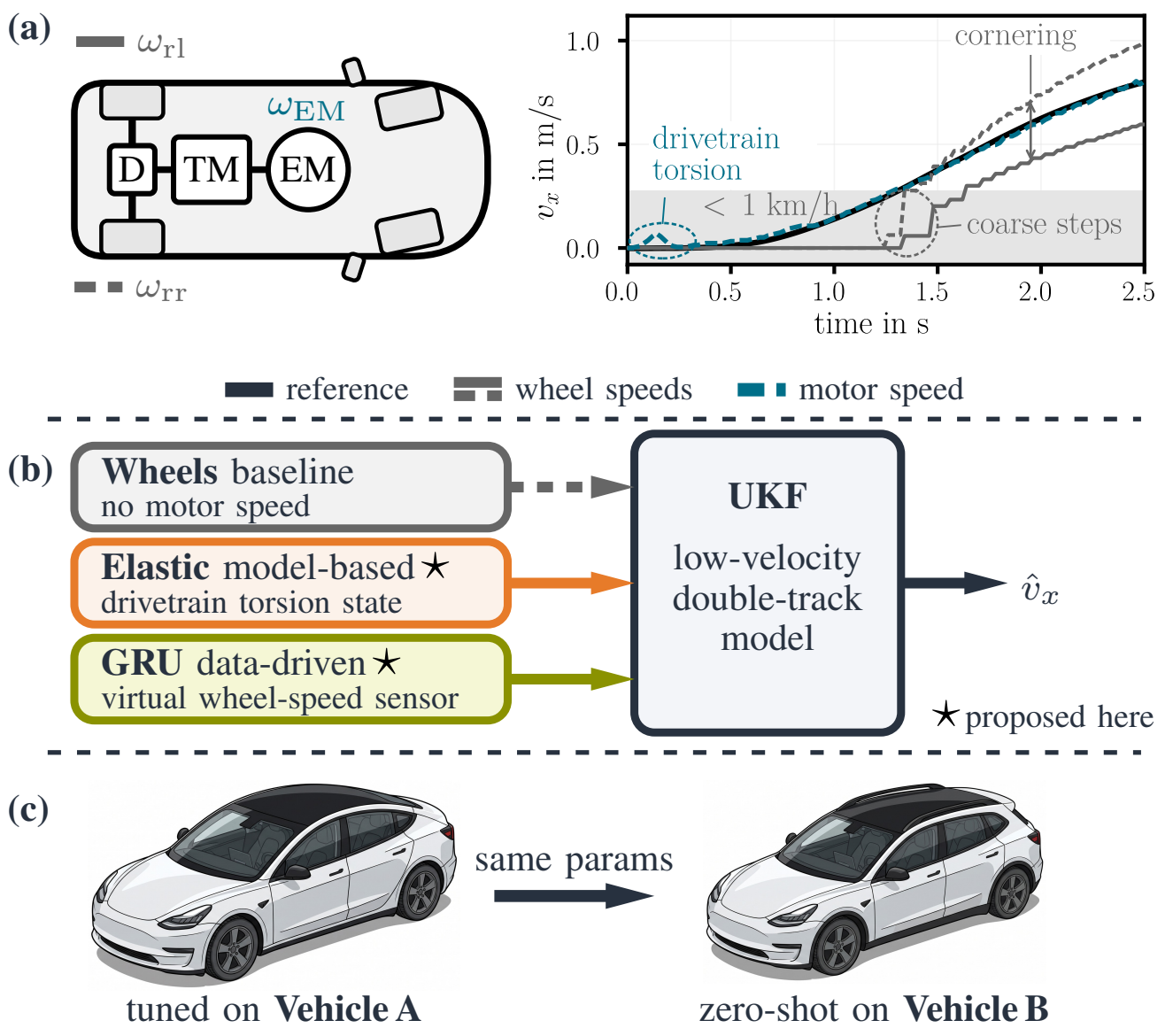}
\caption{Overview of the proposed low-velocity state estimation.
(a)~Source signals of the rear-drive electric drivetrain, the motor speed $\omega_{\mathrm{EM}}$ and the production wheel speeds, of which the panel shows the rear pair $\omega_{\mathrm{rl}}$, $\omega_{\mathrm{rr}}$ (EM electric machine, TM transmission, D differential), on a drive-off from standstill of \VehA{}.
The wheel speeds advance in coarse steps and lag near standstill, and the two rear channels split because the vehicle turns.
The motor speed follows the inertial reference, apart from the marked transient where the driveline winds up before the vehicle breaks away.
(b)~The two proposed motor-speed routes, an explicit drivetrain model (Elastic) and a GRU virtual wheel-speed sensor, feed one double-track UKF and are compared against a wheel-speed-only baseline (Wheels).
(c)~Tuning on \VehA{} transfers zero-shot to \VehB{}, two real series-production vehicles drawn generically for confidentiality.}
\label{fig:concept}
\end{figure}

Longitudinal velocity $v_x$ is a core input for vehicle control, driver-assistance, and automated driving functions, and reviews of vehicle state estimation tie control performance to its accuracy~\cite{guo2018survey,jin2019survey}.
At low speeds, functions such as automated parking and anti-jerk drivetrain control depend on an accurate $v_x$ estimate for precise positioning and ride comfort.

The resolution of a production wheel-speed sensor is set by its fixed number of pulses per revolution, and its update rate by the wheel speed itself.
Below \qty{1}{\kilo\meter\per\hour} both limits become severe, so the reported wheel speed advances in coarse steps and lags the true motion by the time since the last pulse~\cite{ragab2024speed}.
Anti-jerk control in particular suffers from the poor quality or the absence of these signals~\cite{scamarcio2020antijerk}.
Fig.~\ref{fig:concept}(a) shows measurement data from a drive-off from standstill at creep speed.
Once the inertial reference has reached \qty{1}{\kilo\meter\per\hour}, neither rear wheel speed has caught up, with the left one still reading zero and the right one about a fifth of the true speed.
The two rear channels differ by up to \qty{0.4}{\meter\per\second} while the vehicle turns.
Falling back on a vehicle model does not close this resolution gap either, because the model relies on the same wheel speeds and becomes numerically stiff as the velocity decreases~\cite{rill2020dynamics}.
Below \qty{1}{\kilo\meter\per\hour}, the missing velocity information has to come from another sensor.

In an electric vehicle, a second speed source is already on board: the resolver of the electric machine measures the motor speed at a substantially higher resolution than the production wheel-speed sensors, and the signal stays informative down to standstill.
Unlike combustion platforms with a clutch-decoupled driveline, an electric vehicle couples motor and wheels permanently, so the motor speed relates directly to wheel motion.
Fig.~\ref{fig:concept}(a) includes it for the same maneuver, where it follows the reference while the wheel speeds have not yet reacted.
Drivetrain torsion and backlash, however, distort the motor-to-wheel relation.
At the marked pull-away the motor already turns while the vehicle is still standing, so the motor speed is no direct substitute for the degraded wheel speeds.
The transient is small in this gentle drive-off and grows with the transmitted torque.
The maneuver is shown at low torque deliberately, where the degradation of the wheel-speed signal is easiest to see.

The unscented Kalman filter~(UKF) is established for vehicle state estimation~\cite{guo2018survey,jin2019survey,antonov2011ukf,wielitzka2018ukf}.
Its maturity is documented by a production braking system that estimates the vehicle velocity with a UKF~\cite{weber2025abs}.
Reported extensions address four-wheel electric vehicles~\cite{heidefeld2020ukf}, adaptive noise handling with the adaptive UKF (AUKF)~\cite{chen2023sagehusa,alshawi2024adaptive,wang2023adaptive}, and hybrid architectures with model-based virtual measurements~\cite{goblirsch2024race}.
These estimators address normal driving speeds, and none of them integrates the motor speed of the electric drive as a measurement.

At low velocity, the standard slip-based tire description breaks down, so estimators either exclude this regime or switch the true slip kinematics off below a velocity threshold~\cite{heidefeld2020ukf}.
This has motivated low-speed tire models that remove the singularity that the slip definition develops as the wheel speed approaches zero~\cite{bernard1995tire,zhang2018tire,jung2018slip,kim2018slip}.
These models are studied on their own, yet none of them has been used inside a model-based estimator to keep the filter stable below the velocity threshold at which the slip kinematics are otherwise switched off.

One route to exploit the motor speed is to extend the double-track model by the drivetrain that connects the machine to the wheels, so that the torsion becomes part of the model.
Yet stiff torsional springs acting on bodies of small inertia add high eigenfrequencies to the model, and they are the configuration in which an explicit Euler step turns numerically unstable~\cite{arnold2011numerical}.

A double-track model with slip-based tire forces turns numerically stiff at low velocity, regardless of how the motor speed is used.
The gradient of the tire force with respect to slip is steep near zero slip~\cite{rill2020dynamics}, and a lower longitudinal velocity produces eigenvalues with larger real parts in magnitude, so the system grows stiffer as the vehicle slows~\cite{musiu2026benchmark} and the stable step size of an explicit scheme shrinks.
How such a model is discretized therefore decides whether a model-based filter stays stable and accurate within the real-time budget of an on-board implementation, yet the choice has received little attention in the filtering literature.
Numerically stable discrete-time vehicle models and integrators for stiff mechanical systems are available~\cite{rill2006implicit,arnold2011numerical,ge2020stable,zhan2024explicit}, and explicit fixed-step schemes remain the preferred choice in real-time applications because their computational cost is predictable~\cite{musiu2026benchmark}.
Inside UKF estimators, however, the integrator is typically a fixed-step explicit Euler step adopted without justification~\cite{wielitzka2017friction,wielitzka2018ukf,bechtloff2016cornering}.
Recent adaptive UKF variants do not report their discretization at all~\cite{chen2023sagehusa,wang2023adaptive,alshawi2024adaptive}, and the surveys do not treat it~\cite{guo2018survey,jin2019survey}.
Across this literature the discretization is thus neither justified nor compared against alternatives.

Data-driven estimators instead learn the mapping from the on-board sensors to a reference measurement directly~\cite{schaefke2023transformer}, so neither a discretization nor a laborious identification of model parameters is required.
Learning-based virtual wheel-speed sensors apply this idea to the degraded wheel speeds and deliver a high-resolution signal~\cite{schaefke2026virtual}, but they offer neither physical interpretability nor an uncertainty, and they have not been compared against a physics-based drivetrain filter in this regime.
Such networks have also been applied to the longitudinal velocity directly~\cite{annunziata2022longvel,bonfitto2021datadriven}, yet without the drivetrain signal of an electric vehicle and without the sub-\qty{1}{\kilo\meter\per\hour} focus.
Table~\ref{tab:rw_matrix} contrasts the estimators discussed above along the criteria that matter below \qty{1}{\kilo\meter\per\hour}, and each of them covers only part of these criteria.

\begin{table}[!t]
\caption{Positioning Against Related Vehicle State Estimators}
\label{tab:rw_matrix}
\centering
\small
\setlength{\tabcolsep}{3pt}
\renewcommand{\arraystretch}{1.25}
\begin{tabular}{@{}llcccccc@{}}
\toprule
Reference & Filter &
\rot{$<\,$\qty{1}{\kilo\meter\per\hour}~} &
\rot{Slip singularity~} &
\rot{Stiff discret.~} &
\rot{Drivetrain model~} &
\rot{Real vehicle (EV)~} &
\rot{Cross-vehicle~} \\
\midrule
Antonov et al.~\cite{antonov2011ukf}        & UKF     & \tno   & \tno   & \tpart & \tno   & \tpart & \tno \\
Wielitzka et al.~\cite{wielitzka2017friction,wielitzka2018ukf} & UKF & \tno & \tno & \tno & \tno & \tpart & \tno \\
Wielitzka~\cite{wielitzka2020diss}          & UKF     & \tpart & \tno   & \tpart & \tno   & \tpart & \tno \\
Heidfeld et al.~\cite{heidefeld2020ukf}     & UKF     & \tno   & \tyes  & \tpart & \tno   & \tyes  & \tno \\
Chen; Wang~\cite{chen2023sagehusa,wang2023adaptive} & AUKF & \tno & \tno & \tno & \tno & \tno & \tno \\
Pu et al.~\cite{pu2023lowspeed}             & IMM+KF  & \tpart & \tno   & \tno   & \tpart & \tyes  & \tno \\
Alshawi et al.~\cite{alshawi2024adaptive}   & AUKF    & \tno   & \tyes  & \tno   & \tno   & \tyes  & \tno \\
Goblirsch et al.~\cite{goblirsch2024race}   & EKF+UKF & \tno   & \tno   & \tno   & \tno   & \tpart & \tno \\
Sch\"afke et al.~\cite{schaefke2026virtual} & GRU     & \tyes  & \tno   & \tno   & \tpart & \tyes  & \tno \\
\textbf{This work} & \textbf{UKF} & \textbf{\tyes} & \textbf{\tyes} & \textbf{\tyes} & \textbf{\tyes} & \textbf{\tyes} & \textbf{\tyes} \\
\bottomrule
\end{tabular}

\vspace{2pt}
{\footnotesize\raggedright
\tyes~addressed,\quad \tpart~partial or with caveats,\quad \tno~not addressed.\\
\emph{Drivetrain model}: \tyes~motor speed with explicit torsion, \tpart~motor speed only.\\
\emph{Real vehicle (EV)}: \tyes~real EV, \tpart~real non-EV, \tno~simulation only.\par}
\end{table}

Heidfeld et al.~\cite{heidefeld2020ukf} make the numerical exclusion of this regime explicit.
To keep the slip kinematics solvable at a \qty{1}{\milli\second} sample time, they floor the slip velocity at a numerical speed of \qty{3.5}{\meter\per\second}, so below roughly \qty{12.6}{\kilo\meter\per\hour} their filter no longer runs on the true slip kinematics.
The closest model-based effort in the low-speed regime fuses a wheel-speed model and a motor-speed model in an interacting-multiple-model (IMM) estimator, but it does not model the drivetrain torsion that distorts the motor-speed signal~\cite{pu2023lowspeed}.
We are aware of no nonlinear filter that integrates the motor speed for the sub-\qty{1}{\kilo\meter\per\hour} regime, and of no direct comparison of a physical against a learned route that is validated across vehicles.

We estimate $v_x$ with a UKF on a double-track model adapted to low velocities and integrate the motor speed through two routes.
Throughout, we refer to the four configurations by short names.
The first augments the model with an explicit torsional drivetrain, so that the motor speed becomes a state of the estimator (Elastic).
The second uses a gated recurrent unit (GRU) as a virtual wheel-speed sensor and feeds its output into the filter as a pseudo-measurement, which makes this route a hybrid of a model-based and a data-based estimator.
Both routes are compared against the conventional wheel-speed-only configuration (Wheels) and against the baseline that adds the motor-speed sensor to Wheels with the simplest coupling, a rigid driveline without torsion (Rigid).
To keep the comparison fair, all four variants share the integration scheme selected by a systematic discretization study, and each has its process noise tuned by particle-swarm optimization (PSO).
The physical route needs parameters that are laborious to identify, the drivetrain compliance in particular.
The learned route, in turn, is commonly suspected of overfitting to the vehicle and the situations it was trained on.
To test transferability across vehicles, all estimators are tuned on \VehA{} and deployed on a second vehicle, \VehB{}, without any re-tuning, a zero-shot cross-vehicle transfer.
The main contributions of this work are as follows.
\begin{itemize}
\item An explicit torsional drivetrain model inside the UKF, which makes the high-resolution motor speed of an electric vehicle usable for $v_x$ estimation down to standstill.
\item A systematic comparison of explicit and semi-implicit discretization schemes for stiff low-velocity vehicle models inside a state estimator, yielding a real-time fixed-step scheme that is stable down to standstill.
\item A direct comparison of this physics-based route against a hybrid route that feeds the output of a GRU virtual wheel-speed sensor into the same filter. Two baselines show that the gain stems from modeling the torsion, not from the motor-speed sensor alone.
\item A zero-shot cross-vehicle validation in real-world test drives with two series-production electric vehicles: both proposed estimators are tuned on \VehA{}, deployed unchanged on \VehB{}, and retain their advantage in the sub-\qty{1}{\kilo\meter\per\hour} regime.
\end{itemize}
Four experiments, E1 to E4, support these contributions.
They comprise a discretization benchmark against an adaptive-step reference solution (E1), a particle-swarm tuning of the filter noise (E2), the four-way comparison on \VehA{} across low-speed, cornering, and curb-crossing maneuvers (E3), and the zero-shot transfer to \VehB{} (E4).
The remainder of this paper is organized as follows.
Section~\ref{sec:model} develops the low-velocity double-track and drivetrain model, and Section~\ref{sec:discretization} reviews numerical integration and low-speed slip handling and selects the discretization.
The two routes that integrate the motor speed follow in Section~\ref{sec:integration}, the test vehicles and the measurement campaigns in Section~\ref{sec:setup}, and the four-way comparison and the cross-vehicle transfer in Section~\ref{sec:results}.
A discussion and a conclusion close the paper, and Appendix~\ref{app:euler_sweep} adds a closed-loop step-size sweep.

\section{Low-Velocity Vehicle Model}
\label{sec:model}
The estimator rests on a planar double-track model with Magic Formula tire forces and rotational wheel dynamics (Figs.~\ref{fig:doubletrack} and~\ref{fig:tiremodel}).
Section~\ref{subsec:drivetrain} extends this structure by an explicit drivetrain model that couples the electric machine to the driven axle (Fig.~\ref{fig:drivetrain}).
Near standstill, any such slip-based model turns numerically stiff~\cite{rill2020dynamics}, which motivates the discretization study in Section~\ref{sec:discretization}.

\subsection{Planar Double-Track Body Model}
\label{subsec:body}
\begin{figure}[t]
\centering
{\footnotesize\def\svgwidth{\columnwidth}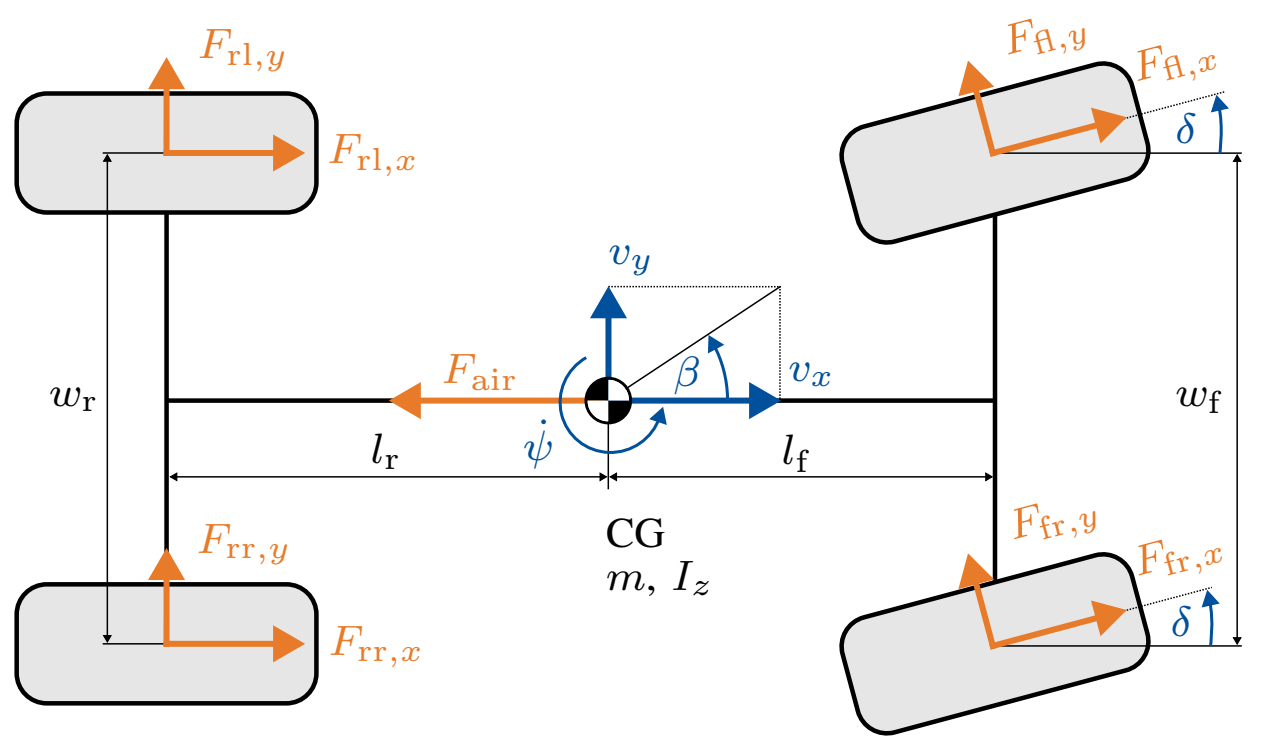}
\caption{Planar double-track vehicle model and its notation: vehicle-fixed frame at the center of gravity (CG) with velocities~$v_x$, $v_y$, side-slip angle~$\beta$, and yaw rate~$\dot\psi$; tire-frame forces~$F_{i,x}$, $F_{i,y}$ at each wheel~$i\in\{\mathrm{fl},\mathrm{fr},\mathrm{rl},\mathrm{rr}\}$; front-wheel steering angle~$\delta$; aerodynamic drag~$F_{\mathrm{air}}$; track widths~$w_{\mathrm{f}}$, $w_{\mathrm{r}}$; CG-to-axle distances~$l_{\mathrm{f}}$, $l_{\mathrm{r}}$; mass~$m$; and yaw inertia~$I_z$.}
\label{fig:doubletrack}
\end{figure}

The chassis is treated as a rigid body with three planar degrees of freedom (Fig.~\ref{fig:doubletrack}), described by the longitudinal velocity $v_x$, lateral velocity $v_y$ and by the yaw rate $\dot\psi$ in a vehicle-fixed frame at the center of gravity (CG).
Each wheel $i\in\{\mathrm{fl},\mathrm{fr},\mathrm{rl},\mathrm{rr}\}$, from front-left to rear-right, has a contact-point velocity in the vehicle frame that combines the body velocity with the yaw-rate contribution of its lever arms.
Superscripts V and T mark quantities expressed in the vehicle coordinate system (CS) $(\mathrm{CS})_{\mathrm{V}}$ and the tire coordinate system $(\mathrm{CS})_{\mathrm{T}}$.
Body quantities are given in $(\mathrm{CS})_{\mathrm{V}}$ and tire quantities in $(\mathrm{CS})_{\mathrm{T}}$, and the superscript is written only where quantities of both frames appear in the same equation.
For the front-left wheel,
\begin{equation}
\label{eq:kinematics}
v^{\mathrm{V}}_{\mathrm{fl},x} = v_x - \dot\psi\,\tfrac{w_{\mathrm{f}}}{2},
\qquad
v^{\mathrm{V}}_{\mathrm{fl},y} = v_y + \dot\psi\,l_{\mathrm{f}},
\end{equation}
with the front and rear track widths~$w_{\mathrm{f}}$, $w_{\mathrm{r}}$ and the CG-to-axle distances~$l_{\mathrm{f}}$, $l_{\mathrm{r}}$; the remaining wheels follow from the same geometry by the corresponding sign and axle.
Rotating by the steering angle $\delta_i$ gives the wheel-center velocity in the tire frame,
\begin{equation}
\label{eq:tireframevel}
\begin{aligned}
v^{\mathrm{T}}_{i,x} &= \phantom{-}v^{\mathrm{V}}_{i,x}\cos\delta_i + v^{\mathrm{V}}_{i,y}\sin\delta_i,\\
v^{\mathrm{T}}_{i,y} &= -v^{\mathrm{V}}_{i,x}\sin\delta_i + v^{\mathrm{V}}_{i,y}\cos\delta_i.
\end{aligned}
\end{equation}
The model carries one steering angle per wheel. The two front angles are derived from the steering-wheel angle with a constant steering ratio and are therefore equal, a common simplification that neglects the Ackermann difference, and the rear angles are zero on both platforms.
The body motion is governed by the planar Newton--Euler equations
\begin{align}
\label{eq:bodyeom}
m\,(\dot v_x - v_y\dot\psi) &= \sum_i F^{\mathrm{V}}_{i,x} - F_{\mathrm{air}},\\
m\,(\dot v_y + v_x\dot\psi) &= \sum_i F^{\mathrm{V}}_{i,y},\notag\\
I_z\,\ddot\psi &= \sum_i M_{z,i},\notag
\end{align}
where $m$ is the vehicle mass, $I_z$ the yaw inertia, $F^{\mathrm{V}}_{i,x}$ and $F^{\mathrm{V}}_{i,y}$ the tire forces projected onto the vehicle frame according to equation~\eqref{eq:frame}, and $F_{\mathrm{air}}$ the aerodynamic drag.
The aerodynamic drag is
\begin{equation}
\label{eq:airdrag}
F_{\mathrm{air}}=\tfrac{1}{2}\rho\,c_{\mathrm{w}}A\,v_x^2\tanh(v_x/v_{\varepsilon}),
\end{equation}
with the air density~$\rho$, the drag coefficient~$c_{\mathrm{w}}$, the frontal area~$A$, and $v_{\varepsilon}=\qty{0.01}{\meter\per\second}$.
Because $v_x^2$ is even, the $\tanh$ factor restores the sign of the drag when the vehicle drives in reverse and keeps it smooth through standstill.
The yaw moment sums the wheel forces over their track and axle lever arms,
\begin{equation}
\label{eq:yawmoment}
\begin{aligned}
\sum_i M_{z,i} =\;& \bigl(F^{\mathrm{V}}_{\mathrm{fr},x}-F^{\mathrm{V}}_{\mathrm{fl},x}\bigr)\tfrac{w_{\mathrm{f}}}{2}
                   + \bigl(F^{\mathrm{V}}_{\mathrm{rr},x}-F^{\mathrm{V}}_{\mathrm{rl},x}\bigr)\tfrac{w_{\mathrm{r}}}{2}\\
                 &+ \bigl(F^{\mathrm{V}}_{\mathrm{fl},y}+F^{\mathrm{V}}_{\mathrm{fr},y}\bigr)l_{\mathrm{f}}
                   - \bigl(F^{\mathrm{V}}_{\mathrm{rl},y}+F^{\mathrm{V}}_{\mathrm{rr},y}\bigr)l_{\mathrm{r}}.
\end{aligned}
\end{equation}
This planar double-track structure is standard in model-based vehicle state estimation~\cite{rill2020dynamics,antonov2011ukf,heidefeld2020ukf}.

\subsection{Tire Model and Low-Velocity Slip Regularization}
\label{subsec:tire}
\begin{figure}[t]
\centering
{\footnotesize\def\svgwidth{\columnwidth}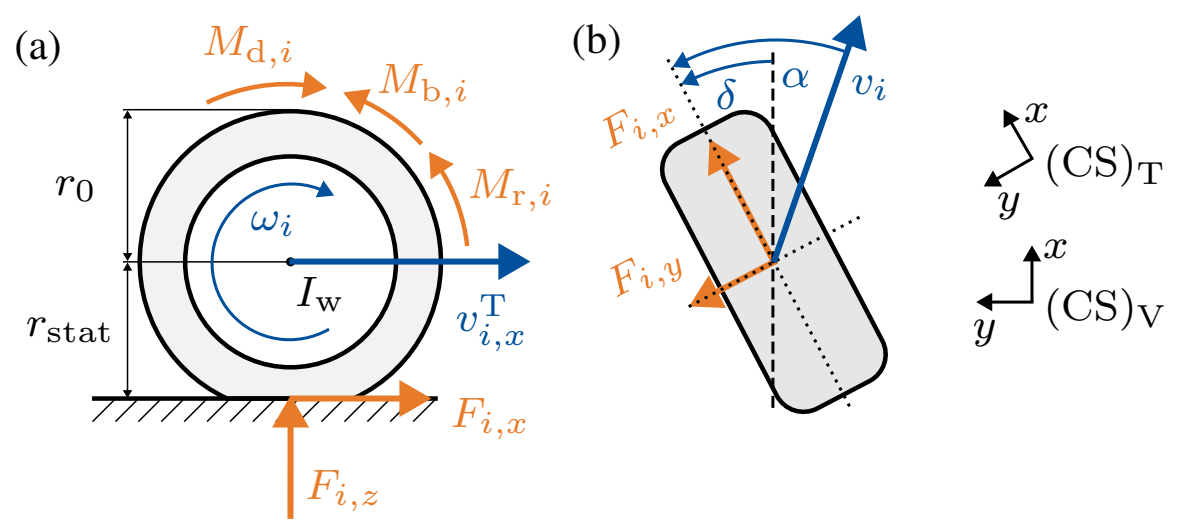}
\caption{Single-wheel tire model. (a)~In-plane wheel dynamics of~\eqref{eq:wheel} with the wheel inertia~$I_{\mathrm{w}}$, the unloaded and static radii~$r_0$, $r_{\mathrm{stat}}$, the wheel-center velocity~$v^{\mathrm{T}}_{i,x}$, and the contact-patch forces~$F_{i,x}$, $F_{i,z}$. (b)~Top view with the tire- and vehicle-fixed coordinate systems, the slip angle~$\alpha$, the wheel-center velocity~$v_i$, and the planar tire forces~$F_{i,x}$, $F_{i,y}$.}
\label{fig:tiremodel}
\end{figure}

The tire model maps the wheel kinematics to contact forces (Fig.~\ref{fig:tiremodel}).
The longitudinal and lateral slip are formed with a regularized transport velocity
\begin{equation}
\label{eq:slip}
\begin{gathered}
v_{\mathrm{tr},i} = \lvert v^{\mathrm{T}}_{i,x}\rvert + v_{\mathrm{reg}},\\[2pt]
s_{x,i} = \frac{\omega_i\,r_{\mathrm{dyn}} - v^{\mathrm{T}}_{i,x}}{v_{\mathrm{tr},i}},
\qquad
s_{y,i} = \frac{-v^{\mathrm{T}}_{i,y}}{v_{\mathrm{tr},i}},
\end{gathered}
\end{equation}
where $\omega_i$ is the wheel speed and $r_{\mathrm{dyn}}$ the dynamic rolling radius.
This radius lies between the static radius $r_{\mathrm{stat}}$ and the unloaded radius $r_0$ of Fig.~\ref{fig:tiremodel}(a) and turns a wheel speed into a rolling velocity.
A slip defined with the wheel-center speed alone is singular at standstill, because the denominator vanishes as $v\rightarrow0$.
The constant $v_{\mathrm{reg}}=\qty{1.0}{\meter\per\second}$ removes this singularity by bounding the denominator away from zero, following the fictitious-velocity regularization of Rill~\cite{rill2020dynamics}.
Alternative low-speed slip formulations are reviewed in Section~\ref{sec:discretization}.
This regularization weakens the modeled tire response at low speed.
At \qty{1}{\kilo\meter\per\hour} it attenuates the modeled slip to about \qty{22}{\percent} of its unregularized value, a model error which the process noise tuned in Section~\ref{subsec:tuning} then absorbs.
Away from standstill, $v_{\mathrm{tr},i}$ approaches $\lvert v^{\mathrm{T}}_{i,x}\rvert$ and both the longitudinal slip and the slip angle recover the standard kinematics of Fig.~\ref{fig:tiremodel}(b).
The combined slip $s_i=\sqrt{s_{x,i}^2+s_{y,i}^2+\varepsilon^2}$, with a small constant $\varepsilon=\num{1e-3}$ that keeps the projection of~\eqref{eq:tireforce} finite at zero slip, drives a simplified Pacejka Magic Formula force~\cite{pacejka1992magic,wielitzka2018ukf} $F^{\mathrm{MF}}_{j}(s_i,F_{i,z})$, evaluated once per force direction $j\in\{x,y\}$, each with its own coefficient set,
\begin{align}
\label{eq:magicformula}
F^{\mathrm{MF}}_{j} &= D\,\sin\!\bigl(C_j\arctan\varphi_j\bigr),\\
\label{eq:mfargument}
\varphi_j &= (1-E_j)\,B_j s_i + E_j\arctan(B_j s_i),
\end{align}
with the dimensionless Magic Formula argument~$\varphi_j$, the peak force~$D=\mu F_{i,z}$, the friction coefficient~$\mu=\num{1.0}$ for dry asphalt, the shape factor~$C_j$, the curvature factor~$E_j$, and the stiffness factor~$B_j=K_j/(C_j D)$ with the load-dependent slip stiffness~$K_j=c_{1,j}\sin\!\bigl(2\arctan(F_{i,z}/c_{2,j})\bigr)$.
The combined force is projected back onto the slip directions,
\begin{equation}
\label{eq:tireforce}
F_{i,x} = F^{\mathrm{MF}}_{x}\,\frac{s_{x,i}}{s_i},
\qquad
F_{i,y} = F^{\mathrm{MF}}_{y}\,\frac{s_{y,i}}{s_i}.
\end{equation}
The tire forces are given in the tire frame.
The vertical load follows from the static axle split, $F_{\mathrm{f},z}$$=m g\,l_{\mathrm{r}}/(2l)$ on the front wheels and $F_{\mathrm{r},z}$$=m g\,l_{\mathrm{f}}/(2l)$ on the rear, with the gravitational acceleration $g$ and the wheelbase $l=l_{\mathrm{f}}+l_{\mathrm{r}}$; the quasi-static load transfer is neglected, since it is minor in the low-velocity regime.
Finally, the tire-frame forces are rotated into the vehicle frame,
\begin{equation}
\label{eq:frame}
\begin{aligned}
F^{\mathrm{V}}_{i,x} &= F_{i,x}\cos\delta_i - F_{i,y}\sin\delta_i,\\
F^{\mathrm{V}}_{i,y} &= F_{i,x}\sin\delta_i + F_{i,y}\cos\delta_i.
\end{aligned}
\end{equation}

\subsection{Wheel Rotational Dynamics}
\label{subsec:wheel}
Each wheel carries a rotational degree of freedom driven by the applied torques,
\begin{equation}
\label{eq:wheel}
I_{\mathrm{w}}\,\dot\omega_i = M_{\mathrm{d},i} - F_{i,x}\,r_{\mathrm{dyn}} - M_{\mathrm{b},i} - M_{\mathrm{r},i},
\end{equation}
with the wheel inertia~$I_{\mathrm{w}}$ and the brake torque~$M_{\mathrm{b},i}$.
The drive torque $M_{\mathrm{d},i}$ is delivered by the drivetrain of Section~\ref{subsec:drivetrain}, which splits the shaft torque equally over the two driven wheels; on the front axle it is zero.
The rolling resistance is modeled as a load-proportional torque with a linear speed dependence~\cite{rill2020dynamics}, $M_{\mathrm{r},i}=F_{i,z}\,(c_{r0}+c_{r1}\lvert\omega_i\rvert)\tanh(\omega_i/\omega_{\varepsilon})$, in which the $\tanh$ factor with $\omega_{\varepsilon}=\qty{1.0}{\radian\per\second}$ acts as a smooth sign function, so the torque opposes the rotation and vanishes continuously at standstill.
The coefficients $c_{r0}$ and $c_{r1}$ act as an equivalent lever arm, so the expression returns a torque without a further radius factor.
The brake torque uses the continuous dry-friction model of Rill~\cite{rill2020dynamics}.
The torque that would arrest the wheel balances the tire force and damps the residual wheel motion,
\begin{equation}
\label{eq:brakearrest}
M_{\mathrm{arr},i} = -F_{i,x}\,r_{\mathrm{dyn}} + r_{\mathrm{dyn}}\sqrt{c_{\mathrm{tx}}I_{\mathrm{w}}}\;\omega_i,
\end{equation}
with the longitudinal tire stiffness~$c_{\mathrm{tx}}=\qty{160}{\kilo\newton\per\meter}$~\cite{rill2020dynamics}.
The applied brake torque is the arresting torque $M_{\mathrm{arr},i}$, capped at the commanded brake torque, so that a locked wheel is held at standstill instead of being driven backwards by the tire force.

\subsection{Drivetrain Extension: Two-Mass Oscillator}
\label{subsec:drivetrain}
\begin{figure}[t]
\centering
{\footnotesize\def\svgwidth{\columnwidth}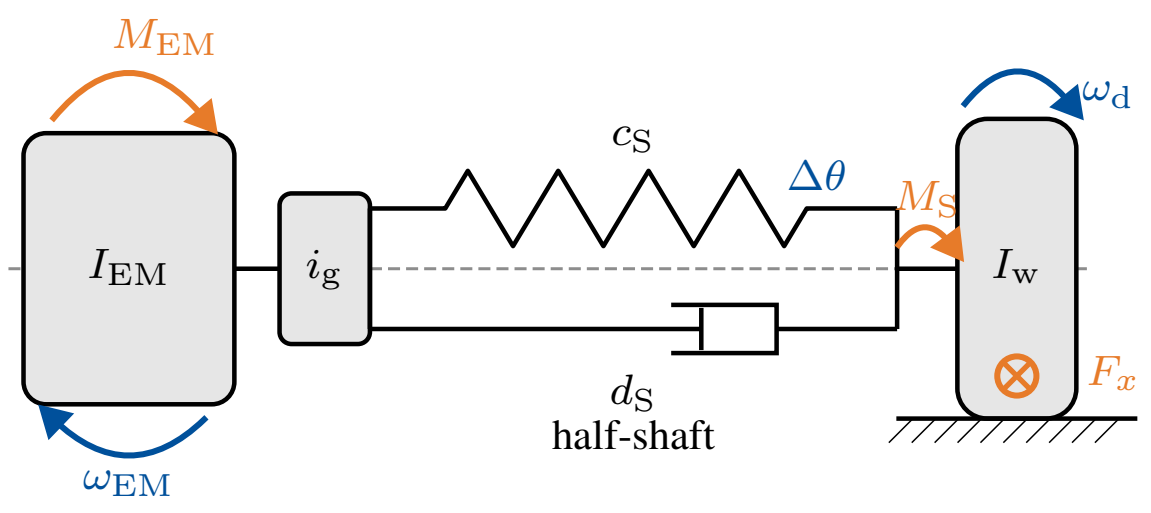}
\caption{Drivetrain as a two-mass oscillator (the proposed Elastic extension), viewed along the vehicle's longitudinal axis.
The electric motor with inertia~$I_{\mathrm{EM}}$, driven by the torque~$M_{\mathrm{EM}}$ and turning at~$\omega_{\mathrm{EM}}$, couples through the gear ratio~$i_{\mathrm{g}}$ and a compliant half-shaft to the driven wheels (inertia~$I_{\mathrm{w}}$, mean speed~$\omega_{\mathrm{d}}$), one of which is drawn.
The half-shaft is a torsional spring~$c_{\mathrm{S}}$ and damper~$d_{\mathrm{S}}$ in parallel, and the twist~$\Delta\theta$ and the transmitted shaft torque~$M_{\mathrm{S}}$ couple the two inertias~\eqref{eq:driveline}.
The longitudinal tire force~$F_x$ of~\eqref{eq:wheel} acts at the contact patch and points into the drawing plane.
In the rigid limit the shaft is stiff, so $\omega_{\mathrm{EM}}=i_{\mathrm{g}}\omega_{\mathrm{d}}$ and $\Delta\theta=0$.}
\label{fig:drivetrain}
\end{figure}

On an electric vehicle the motor speed is available at high resolution, but drivetrain compliance and backlash make it deviate measurably from the wheel speeds whenever the load changes.
A model of this coupling is required for the motor speed to inform the longitudinal velocity.
This section derives a compliant driveline and its rigid limit, and Section~\ref{subsec:variants} builds estimator variants on both.
The half-shaft is modeled as a torsional spring-damper, a two-mass oscillator between the motor and the driven axle (Fig.~\ref{fig:drivetrain}),
\begin{equation}
\label{eq:driveline}
\begin{aligned}
M_{\mathrm{S}} &= c_{\mathrm{S}}\,\Delta\theta + d_{\mathrm{S}}\,\Delta\dot\theta,\\
I_{\mathrm{EM}}\,\dot\omega_{\mathrm{EM}} &= M_{\mathrm{EM}} - M_{\mathrm{S}}/i_{\mathrm{g}},\\
\Delta\dot\theta &= \omega_{\mathrm{EM}}/i_{\mathrm{g}} - \omega_{\mathrm{d}},
\end{aligned}
\end{equation}
with the motor inertia~$I_{\mathrm{EM}}$, motor speed~$\omega_{\mathrm{EM}}$, motor torque~$M_{\mathrm{EM}}$, gear ratio~$i_{\mathrm{g}}$, shaft stiffness~$c_{\mathrm{S}}$, shaft damping~$d_{\mathrm{S}}$, torsion angle~$\Delta\theta$, and shaft torque~$M_{\mathrm{S}}$.
The driven-axle speed $\omega_{\mathrm{d}}$ is the mean speed of the driven wheels, and $M_{\mathrm{S}}$ is split equally as the drive torque $M_{\mathrm{d},i}$ over those wheels.
The rigid driveline is the stiff-shaft limit, in which the driveline states are kinematically slaved, $\omega_{\mathrm{EM}}=i_{\mathrm{g}}\omega_{\mathrm{d}}$ and $\Delta\theta=0$, and the gear-scaled motor torque passes straight to the driven axle.
In this limit the gear-scaled motor inertia is reflected onto the driven axle and distributed equally over the two driven wheels, whose inertia in~\eqref{eq:wheel} becomes
\begin{equation}
\label{eq:inertia_rigid}
I^{\mathrm{eff}}_{\mathrm{w},i} = I_{\mathrm{w}} + \tfrac{1}{2}\,i_{\mathrm{g}}^{2}\,I_{\mathrm{EM}},
\end{equation}
whereas the compliant driveline carries $I_{\mathrm{EM}}$ as a state of its own.
This lumped form is exact for the common mode of the driven axle, the mode the motor speed follows.
The compliant shaft introduces a torsional mode at a higher frequency than the body and wheel dynamics, a second source of numerical stiffness on top of the tire-force gradient.
This combined numerical stiffness is what the discretization study in Section~\ref{sec:discretization} must handle.
This half-shaft spring-damper follows Rill~\cite{rill2020dynamics}.

\subsection{Continuous-Time Process and Measurement Model}
\label{subsec:plant}
The states and inputs of the model are collected in
\begin{equation}
\label{eq:state}
\mm{x} = \bigl[\dot\psi,\,v_x,\,v_y,\,\omega_{\mathrm{fl}},\,\omega_{\mathrm{fr}},\,\omega_{\mathrm{rl}},\,\omega_{\mathrm{rr}},\,\omega_{\mathrm{EM}},\,\Delta\theta\bigr]\transpose \in \R^{9},
\end{equation}
\begin{equation}
\label{eq:input}
\mm{u} = \bigl[\delta_{\mathrm{fl}},\dots,\delta_{\mathrm{rr}},\,M_{\mathrm{EM}},\,M_{\mathrm{b},\mathrm{fl}},\dots,M_{\mathrm{b},\mathrm{rr}}\bigr]\transpose \in \R^{9}.
\end{equation}
The state comprises the yaw rate, the body velocities, the four wheel speeds, the motor speed, and the torsion angle; the input comprises the per-wheel steering angles, the motor torque, and the per-wheel brake torques.
The equations above define the continuous-time process model $\dot{\mm{x}}=\mm{f}(\mm{x},\mm{u})$.
The measured quantities are the yaw rate, the body accelerations, the four wheel speeds, and the motor speed, collected in the measurement vector
\begin{equation}
\label{eq:meas}
\mm{z} = \bigl[\dot\psi,\,a_x,\,a_y,\,\omega_{\mathrm{fl}},\dots,\omega_{\mathrm{rr}},\,\omega_{\mathrm{EM}}\bigr]\transpose {}\in \R^{8}.
\end{equation}
The measurement model $\mm{z}=\mm{h}(\mm{x},\mm{u})$ maps the state onto these quantities; the accelerations follow from the tire forces of the process model as $a_x=\bigl(\sum_i F^{\mathrm{V}}_{i,x}-F_{\mathrm{air}}\bigr)/m$ and $a_y=\sum_i F^{\mathrm{V}}_{i,y}/m$.
A global navigation satellite system (GNSS) ground-speed measurement would be a natural addition to~\eqref{eq:meas}, but GNSS is unavailable or degraded in environments such as underground parking garages and tunnels, which is where target maneuvers of this work take place.
The measurement vector relies on standard on-board sensors only.
Section~\ref{subsec:variants} defines four estimator variants that differ in whether the rigid or the compliant form of $\mm{f}$ is used and in whether and how $\omega_{\mathrm{EM}}$ enters $\mm{h}$.

\section{Discretization Study}
\label{sec:discretization}
The estimator propagates the continuous-time model $\dot{\mm{x}}=\mm{f}(\mm{x},\mm{u})$ of Section~\ref{sec:model} across one sampling interval of $T_{\mathrm{s}}=\qty{20}{\milli\second}$ for every sigma point of the unscented transform (Section~\ref{subsec:filter}).
The discretization scheme that performs this propagation thus determines the stability and the accuracy of the filter, and a large share of its computational cost.
A filter on embedded hardware further requires a fixed number of operations per interval, so adaptive step-size control is ruled out in the filter loop and adaptive stiff solvers can serve only as a reference.
This section benchmarks fixed-step schemes and sub-step budgets and selects the configuration used in all later experiments.

Inside model-based vehicle state estimators, this discretization step receives little attention.
Wielitzka et al.~\cite{wielitzka2017friction,wielitzka2018ukf} and Bechtoff et al.~\cite{bechtloff2016cornering} discretize their models with a first-order explicit Euler step without discussing stability or alternatives.
Antonov et al.~\cite{antonov2011ukf} state their scheme, a Taylor--Lie series expansion, and raise its order for the faster wheel states, but remain within the explicit family and compare no alternatives.
Recent adaptive UKF variants do not report their discretization at all~\cite{chen2023sagehusa,wang2023adaptive,alshawi2024adaptive}.
Stiffness-aware integration of vehicle models is established outside the filtering literature.
Rill~\cite{rill2006implicit} achieves real-time simulation of stiff vehicle dynamics with a modified implicit Euler algorithm, Wielitzka~\cite{wielitzka2020diss} applies a partial-implicit Euler step with a single Newton iteration to the stiff wheel dynamics of a vehicle simulation model, and Musiu et al.~\cite{musiu2026benchmark} analyze the stability limits of explicit fixed-step integration for real-time model predictive control.
To our knowledge, no prior work compares discretization schemes and sub-step budgets systematically inside a vehicle state estimator.

A complementary line of work removes the low-speed slip singularity at the model level.
Rill~\cite{rill2020dynamics} bounds the slip denominator with a small fictitious velocity and notes the resulting trade-off: small values leave the wheel dynamics stiff, and large values degrade accuracy.
Bernard and Clover~\cite{bernard1995tire} promote both slip quantities to first-order lag states with relaxation-length dynamics, which behave well at standstill but add states and require auxiliary damping near zero speed.
Jung et al.~\cite{jung2018slip} and Kim et al.~\cite{kim2018slip} derive the denominator floor from an eigenvalue analysis that ties it to the step size of the explicit integrator, which makes the remedy integrator-aware.
Ge et al.~\cite{ge2020stable} and Zhan et al.~\cite{zhan2024explicit} restructure the discrete-time model itself: a backward-Euler-inspired update of the velocity states with a linear tire model yields an explicit form whose denominators stay bounded as $v\rightarrow0$.
These remedies act on the model or on single integrator mechanisms, and none of them has been evaluated in combination with a state estimator.
The model of Section~\ref{subsec:tire} adopts the regularized denominator of~\eqref{eq:slip}.
The following benchmark answers which fixed-step scheme should propagate the resulting, still stiff dynamics, and at what sub-step budget.

The benchmark compares three fixed-step candidates: the explicit Euler scheme, the classical fourth-order Runge--Kutta scheme (RK4), and a semi-implicit Euler scheme that solves the stiff wheel block with a single Newton step per sub-step and treats the remaining states explicitly.
Each scheme divides the sampling interval into $n$ equal sub-steps, with $n$ swept between 1 and 150 on scheme-dependent grids.
The test case is a \qty{50}{\second} open-loop simulation with synthetic drive-torque, steering, and brake inputs that combine acceleration, sinusoidal steering, and braking with repeated pull-away and brake-to-standstill cycles, so that it covers the excitation patterns of the target regime.
All schemes integrate the same right-hand side $\mm{f}$ of the process model $\dot{\mm{x}}=\mm{f}(\mm{x},\mm{u})$.
The reference is that same model again, integrated by the adaptive eighth-order explicit Runge--Kutta solver~\texttt{Dopri8}~\cite{prince1981highorder} at tolerances $\mathrm{rtol}=10^{-10}$ and $\mathrm{atol}=10^{-12}$, which is free to adapt its step size because it need not run in real time.
Each fixed-step solution is then scored by the RMSE of $v_x$ against this reference.
The measured error is therefore the integration error alone and is not confounded by model mismatch.
Because a biased reference would distort the ranking, the reference is cross-checked against an independent implicit stiff solver (\texttt{Kvaerno5}, a diagonally implicit Runge--Kutta scheme~\cite{kvaerno2004singly}); the deviation is \qty{0.06}{\percent} of the smallest measured discretization error.

\begin{figure}[t]
\centering
\includegraphics{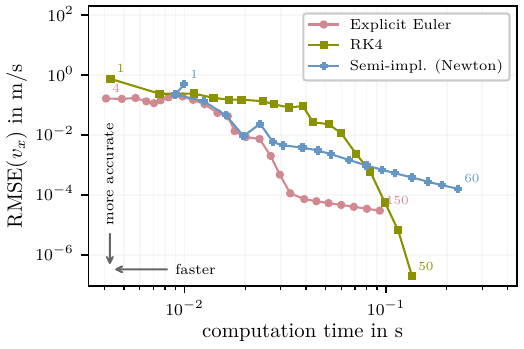}
\caption{Discretization benchmark (E1) on the combined low-speed maneuver as a work-precision diagram: RMSE of $v_x$ against the adaptive reference versus wall-clock computation time, both on logarithmic axes; accuracy and cost improve toward the lower left. Each marker is one sub-step count; the labels give the coarsest and finest count of each scheme.}
\label{fig:discretization}
\end{figure}

Fig.~\ref{fig:discretization} summarizes the outcome as a work-precision diagram: integration error versus wall-clock cost, both on logarithmic axes, with one marker per sub-step count.
Every swept configuration stays stable in this open-loop test, and the schemes differ in accuracy per cost.
Cost is expressed as the real-time factor (RTF), the ratio of simulated time to the wall-clock time of the integration, measured on a laptop processor (Apple M5 Pro).
On the \qty{50}{\second} maneuver an RTF of \num{500} therefore corresponds to a plotted computation time of \qty{0.1}{\second}.
Absolute values are hardware-specific, whereas the ranking at a given budget rests on the relative per-step costs of the schemes.
With a generous compute budget (RTF $\lesssim 450$), RK4 dominates and reaches an RMSE of \qty{1.9e-7}{\meter\per\second} at $n=50$ (RTF $\approx 350$).
At tighter budgets (RTF $\gtrsim 450$) the explicit Euler scheme is the most accurate choice.
The semi-implicit scheme is never the accuracy winner on this model.
At RTF $\approx 500$ it reaches \qty{7.0e-4}{\meter\per\second}, more than an order of magnitude above both explicit schemes at the same cost, so its stability advantage does not pay off once the regularized slip has bounded the tire stiffness.
The deployed configuration of this work, the explicit Euler scheme with $n=100$ sub-steps of \qty{0.2}{\milli\second}, operates at RTF $\approx\numrange{750}{800}$ with an integration error near \qty{4.4e-5}{\meter\per\second}, three orders of magnitude below the estimation errors reported in Section~\ref{sec:results}.

We repeated the study with the elastic-driveline model of Section~\ref{subsec:drivetrain}, using the identified stiffness parameters and the physical gear ratio.
The deployment-relevant conclusions are unchanged: the explicit Euler scheme remains the most accurate at the deployment budget and the crossover to RK4 stays below it, whereas at sub-step budgets far coarser than the deployment operating point the semi-implicit scheme becomes the most accurate, consistent with the numerical stiffness that the compliant drivetrain adds.
All subsequent experiments, tuning and deployment alike, therefore integrate with the explicit Euler scheme at $n=100$ sub-steps.
This yields a discrete-time model that stays stable to standstill at real-time cost.
Appendix~\ref{app:euler_sweep} confirms this stability in closed loop inside the running filter.

\section{Integrating Motor Speed Into Longitudinal Velocity Estimation}
\label{sec:integration}
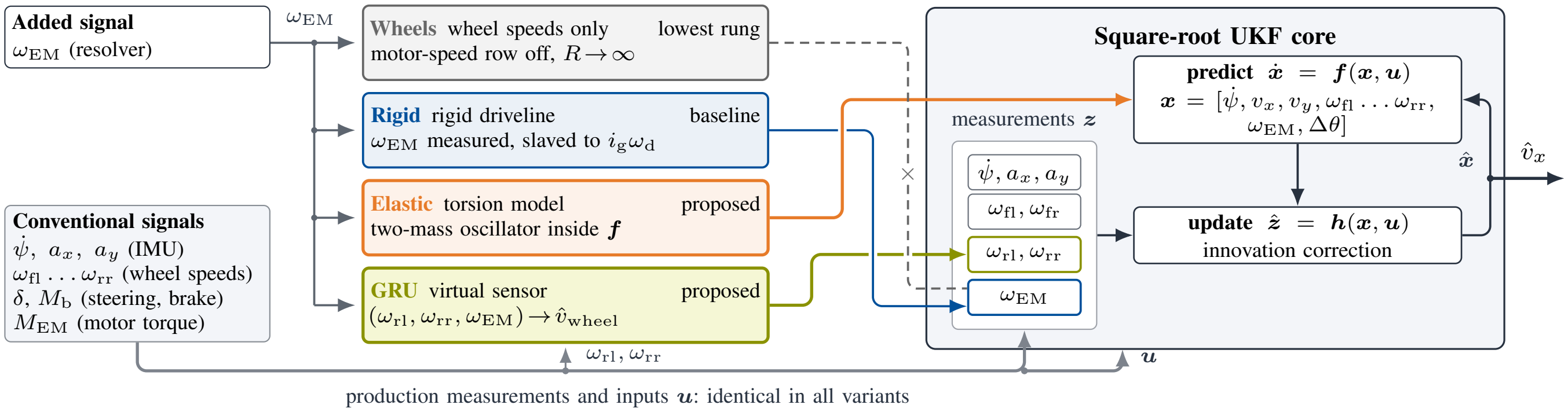
\begin{figure*}[!t]
\centering
\begin{tikzpicture}[
  x=1cm, y=1cm,
  font=\footnotesize,
  >={Latex[length=2mm,width=1.6mm]},
  grp/.style   ={rounded corners=2pt, draw=cInk!55, line width=0.4pt, fill=cFill,
                 align=left, inner sep=2.5pt, font=\scriptsize, text width=2.85cm},
  branch/.style={rounded corners=2pt, draw, line width=0.7pt, align=left,
                 text width=4.45cm, inner sep=2.5pt, minimum height=0.85cm,
                 font=\scriptsize},
  kstep/.style ={rounded corners=2pt, draw=cInk, line width=0.5pt, fill=white,
                 align=center, inner sep=2pt, font=\scriptsize, text width=3.60cm},
  tile/.style  ={rounded corners=1.5pt, draw=cInk!70, line width=0.4pt, fill=white,
                 align=center, inner sep=1.2pt, font=\scriptsize, text width=1.20cm,
                 minimum height=0.40cm},
  flow/.style  ={-{Latex[length=2mm,width=1.6mm]}, line width=0.7pt, rounded corners=3pt},
  trunk/.style ={-{Latex[length=2mm,width=1.6mm]}, line width=1.0pt, draw=cInk!60,
                 rounded corners=3pt},
  feed/.style  ={draw=cInk!60, line width=0.6pt, rounded corners=3pt},
  hop/.style   ={draw=white, line width=2.8pt},
]

\def\yW{4.55}\def\yR{3.55}\def\yE{2.55}\def\yG{1.55}

\node[branch, draw=cWheels, fill=cWheels!8] (vW) at (6.49,\yW) {%
  \textbf{\textcolor{cWheels}{Wheels}}\, wheel speeds only \hfill lowest rung\\
  motor-speed row off, $R\!\to\!\infty$};
\node[branch, draw=cRigid, fill=cRigid!8] (vR) at (6.49,\yR) {%
  \textbf{\textcolor{cRigid}{Rigid}}\, rigid driveline \hfill baseline\\
  $\omega_{\mathrm{EM}}$ measured, slaved to $i_{\mathrm{g}}\omega_{\mathrm{d}}$};
\node[branch, draw=cElastic, fill=cElastic!10, line width=1.0pt] (vE) at (6.49,\yE) {%
  \textbf{\textcolor{cElastic}{Elastic}}\, torsion model \hfill proposed\\
  two-mass oscillator inside $\mm{f}$};
\node[branch, draw=cGRUink, fill=cGRU!18, line width=1.0pt] (vG) at (6.49,\yG) {%
  \textbf{\textcolor{cGRUink}{GRU}}\, virtual sensor \hfill proposed\\
  $(\omega_{\mathrm{rl}},\omega_{\mathrm{rr}},\omega_{\mathrm{EM}})\!\to\!\hat v_{\mathrm{wheel}}$};

\node[grp, fill=white, draw=cInk, anchor=north] (sd) at (1.60,4.975) {\textbf{Added signal}\\[1pt]
  $\omega_{\mathrm{EM}}$ (resolver)};
\node[grp, anchor=south] (sv) at (1.60,1.125) {\textbf{Conventional signals}\\[1pt]
  $\dot\psi,\ a_x,\ a_y$ (IMU)\\ $\omega_{\mathrm{fl}}\dots\omega_{\mathrm{rr}}$ (wheel speeds)\\
  $\delta$, $M_{\mathrm{b}}$ (steering, brake)\\ $M_{\mathrm{EM}}$ (motor torque)};

\draw[flow, cInk!75] (sd.east |- 0,\yW) -- (4.09,\yW);
\draw[cInk!75, line width=0.7pt] (3.62,\yW) -- (3.62,\yG);
\foreach \y in {\yR,\yE,\yG} \draw[flow, cInk!75] (3.62,\y) -- (4.09,\y);
\foreach \y in {\yW,\yR,\yE} \fill[cInk!75] (3.62,\y) circle (1.0pt);
\node[font=\scriptsize, text=cInk, anchor=south] at (3.58,4.63) {$\omega_{\mathrm{EM}}$};

\node[rounded corners=3pt, draw=cInk, line width=0.6pt, fill=cFill,
      minimum width=6.60cm, minimum height=3.90cm] (core) at (13.91,3.00) {};
\node[anchor=north] at (core.north) [yshift=-3pt] {\textbf{Square-root UKF core}};

\node[rounded corners=2pt, draw=cInk!45, line width=0.4pt, fill=white,
      minimum width=1.65cm, minimum height=2.15cm] (zbox) at (11.73,2.35) {};
\node[font=\scriptsize, text=cInk, anchor=south] at (11.73,3.48) {measurements $\mm{z}$};
\node[tile] (t1) at (11.73,3.07) {$\dot\psi, a_x, a_y$};
\node[tile] (t2) at (11.73,2.62) {$\omega_{\mathrm{fl}},\omega_{\mathrm{fr}}$};
\node[tile, draw=cGRUink, line width=0.7pt] (t3) at (11.73,2.13) {$\omega_{\mathrm{rl}},\omega_{\mathrm{rr}}$};
\node[tile, draw=cRigid, line width=0.7pt] (t4) at (11.73,1.64) {$\omega_{\mathrm{EM}}$};

\node[kstep] (pred) at (14.85,3.90)
  {\textbf{predict}\ \ $\dot{\mm{x}}=\mm{f}(\mm{x},\mm{u})$\\[1pt]
   $\mm{x}=[\dot\psi, v_x, v_y, \omega_{\mathrm{fl}}\dots\omega_{\mathrm{rr}},$\\
   $\omega_{\mathrm{EM}}, \Delta\theta]$};
\node[kstep] (upd) at (14.85,2.35)
  {\textbf{update}\ \ $\hat{\mm{z}}=\mm{h}(\mm{x},\mm{u})$\\[1pt] innovation correction};
\draw[flow, cInk] (pred.south) -- (upd.north);
\draw[flow, cInk] (upd.east) -- ++(0.32,0) coordinate (fbx) |- (pred.east)
  node[pos=0.28, anchor=east, font=\scriptsize, text=cInk, xshift=-1.5pt] {$\hat{\mm{x}}$};
\draw[flow, cInk] (zbox.east) -- (upd.west);

\fill[cInk] (fbx |- 0,3.00) circle (1.1pt);
\draw[flow, line width=1.0pt, cInk] (fbx |- 0,3.00) -- (17.90,3.00);
\node[anchor=south] at (17.55,3.06) {$\hat v_x$};

\draw[trunk] (sv.south) -- (1.60,0.80) -- (11.73,0.80) -- (zbox.south);
\draw[trunk] (11.73,0.80) -- (12.85,0.80) -- (12.85,1.05);
\node[font=\scriptsize, text=cInk, anchor=west] at (12.93,0.95) {$\mm{u}$};
\foreach \p in {(6.49,0.80),(11.73,0.80)} \fill[cInk!60] \p circle (1.1pt);
\draw[feed, {-Latex[length=1.8mm,width=1.4mm]}] (6.49,0.80) -- ([yshift=-1pt]vG.south);
\node[font=\scriptsize, text=cInk, anchor=west] at (6.60,0.97) {$\omega_{\mathrm{rl}},\omega_{\mathrm{rr}}$};
\node[font=\scriptsize, text=cInk, anchor=north] at (7.20,0.73)
  {production measurements and inputs $\mm{u}$: identical in all variants};

\draw[flow, cGRUink, line width=1.0pt] (vG.east) -- (9.30,\yG) -- (9.30,2.13) -- (t3.west);
\draw[hop] (10.00,2.00) -- (10.00,2.26);
\draw[flow, cRigid] (vR.east) -- (10.00,\yR) -- (10.00,1.54) -- ([yshift=-2.9pt]t4.west);
\draw[hop] (9.65,3.42) -- (9.65,3.68);
\draw[flow, cElastic, line width=1.0pt] (vE.east) -- (9.65,\yE) -- (9.65,3.90) -- (pred.west);
\draw[hop] (10.40,2.00) -- (10.40,2.26);
\draw[cWheels, densely dashed, line width=0.7pt, rounded corners=3pt]
  (vW.east) -- (10.40,\yW) -- (10.40,1.74) -- ([yshift=+2.9pt]t4.west);
\node[cWheels, font=\scriptsize\bfseries, fill=white, inner sep=0.5pt] at (10.40,3.05) {$\times$};

\end{tikzpicture}
\caption{Hybrid UKF architecture.
All four variants share the trunk of production measurements and inputs $\mm{u}$ (bottom) and the same square-root UKF core with its unscented predict and update steps.
The tiles of $\mm{z}$ group its rows as in \eqref{eq:meas}, with the rear wheel speeds in rows 6 and 7 and the motor speed in row 8.
Only the motor speed $\omega_{\mathrm{EM}}$ is routed through the four colored branches, and each branch ends where it acts on the estimator.
Wheels switches row 8 off ($R\!\to\!\infty$), Rigid corrects that row directly while the rigid driveline slaves $\omega_{\mathrm{EM}}$ to $i_{\mathrm{g}}\omega_{\mathrm{d}}$, the proposed Elastic augments the process model $\mm{f}$ with the two-mass drivetrain oscillator, and GRU writes the output of the learned virtual sensor into rows 6 and 7.
The variant colors match the result figures.}
\label{fig:ukf_architecture}
\end{figure*}

The model of Section~\ref{sec:model} admits several ways of turning the motor speed into information about $v_x$.
This section defines the four estimator variants that this work compares.
They share one filter core, one measurement vector, and one discretization scheme, and they differ only in how $\omega_{\mathrm{EM}}$ reaches the state estimate (Fig.~\ref{fig:ukf_architecture}).

\subsection{Square-Root Unscented Kalman Filter Core}
\label{subsec:filter}
All four variants estimate the state with the same unscented Kalman filter (UKF).
The unscented transform propagates \mbox{$2n_x+1$} deterministic sigma points, with the state dimension~$n_x$ of~\eqref{eq:state}, through the nonlinear model and recovers the state mean and covariance from the propagated points, which avoids the Jacobian linearization of the extended Kalman filter~\cite{julier1997ukf}.
This matters near standstill, where the regularized slip of~\eqref{eq:slip} and the Magic Formula of~\eqref{eq:magicformula} make the right-hand side nonlinear.
The sigma points use the standard scaling $\alpha=10^{-3}$, $\beta=2$, and $\kappa=0$.

The implementation is a square-root UKF~\cite{vandermerwe2001srukf}.
It carries a Cholesky factor of the state covariance instead of the covariance itself and updates that factor through a QR decomposition of the weighted sigma deviations followed by a rank-one Cholesky update.
The covariance therefore stays positive semidefinite by construction, without refactorizing it at every step.
Near standstill the wheel and drivetrain states are far better observed than the body velocities, so the covariance becomes ill-conditioned and a conventional formulation can lose its factorization.

Each predict step integrates every sigma point across one sampling interval of $T_{\mathrm{s}}=\qty{20}{\milli\second}$ with the explicit Euler scheme at $n=100$ sub-steps, the configuration selected in Section~\ref{sec:discretization}.
The update step then corrects the estimate with the measurement vector of~\eqref{eq:meas}.

\subsection{Four Estimator Variants}
\label{subsec:variants}
The variants form an ablation ladder that separates the effect of the added sensor from that of the drivetrain model.

The wheel-speed-only double-track UKF (\textbf{Wheels}) runs on the rigid form of the process model and uses production wheel speeds only.
The motor-speed row of~\eqref{eq:meas} is disabled by inflating its measurement noise, so the filter receives no drivetrain information at all.
Wheels is the lowest rung of the ladder and represents the conventional estimator that an electric drivetrain signal could improve on.

The rigid-driveline UKF (\textbf{Rigid}) enables that motor-speed row, so the state $\omega_{\mathrm{EM}}$ is corrected directly, and the rigid driveline of Section~\ref{subsec:drivetrain} ties it to $i_{\mathrm{g}}\omega_{\mathrm{d}}$.
The torsion angle $\Delta\theta$ of~\eqref{eq:state} stays in the state vector of both rigid variants, but it is inert there, slaved kinematically and entering neither $\mm{f}$ nor $\mm{h}$.
The process model is unchanged, so whatever separates Rigid from Wheels is the value of the extra sensor alone.
Rigid serves as the baseline against which the two proposed variants are reported, and the near-coincidence of Wheels and Rigid in Section~\ref{sec:results} shows that this baseline is not a weak one.

The proposed elastic-driveline UKF (\textbf{Elastic}) instead runs on the compliant process model of~\eqref{eq:driveline}.
Its state carries the motor speed $\omega_{\mathrm{EM}}$ and the torsion angle $\Delta\theta$, the motor torque acts as an input, and the measured motor speed corrects the corresponding state.
The half-shaft twist is therefore estimated, and the motor speed informs $v_x$ through the physical torque path.

The GRU virtual-sensor UKF (\textbf{GRU}) is the hybrid of a model-based and a data-based component.
It keeps the rigid process model and the motor-speed measurement of Rigid and replaces the two rear production wheel-speed measurements with the output of a learned virtual sensor, described next.

\subsection{GRU Virtual Wheel-Speed Sensor}
\label{subsec:gru}
The virtual sensor is a single-layer GRU with 64 hidden units and a linear output head, mapping on-board signals to the velocities of the two driven wheels.
Sch\"afke et al.~\cite{schaefke2026virtual} introduced this sensor and validated it against an inertial reference, where it removed a large part of the low-velocity error of the production wheel-speed signal and generalized across maneuvers.

The deployed network takes three inputs: the two rear production wheel speeds and the motor speed.
The final model of the prior work takes five inputs, adding the drive and brake torque, and was marginally more accurate on the vehicle it was trained on.
Torque inputs are vehicle-specific, however, since the same motor torque produces a different acceleration on a vehicle of different mass and gear ratio.
The reduced input set is therefore chosen deliberately, to keep the sensor transferable to a second vehicle.

The network output is a linear velocity per rear wheel.
It is converted to a wheel rate with the dynamic rolling radius and written into the two rear wheel-speed rows of~\eqref{eq:meas}.
The measurement function $\mm{h}$, the measurement noise $\mm{R}$, and the filter itself are unchanged; only the source of these two measurements changes.
GRU builds on Rigid, so the motor speed remains a model state and its measurement row is corrected as before.
The network is trained with a mean-squared-error loss on \VehA{} recordings that are disjoint from the five held-out test recordings of Section~\ref{subsec:maneuvers}. It is then applied unchanged on \VehB{}, in the same zero-shot manner as the physical parameters of Section~\ref{subsec:vehicles}.

\subsection{Parameter Tuning}
\label{subsec:tuning}
The process noise $\mm{Q}$ is tuned per variant on the training recordings defined in Section~\ref{subsec:maneuvers}, while the measurement noise $\mm{R}$ is fixed a priori from sensor specifications and held identical across variants.
The objective is the mean over the training recordings of the RMSE of the estimated velocity $\hat v_x$ against the ADMA reference $v_x$,
\begin{equation}
\label{eq:pso_objective}
J(\mm{Q}) = \frac{1}{M}\sum_{m=1}^{M}\sqrt{\frac{1}{\lvert\mathcal{K}_m\rvert}\sum_{k\in\mathcal{K}_m}\bigl(\hat v_{x,k}^{(m)}-v_{x,k}^{(m)}\bigr)^2},
\end{equation}
with the $M$ training recordings and the scored samples~$\mathcal{K}_m$ of each.
A candidate that diverges on a recording is scored at a fixed penalty, so it cannot poison the swarm.
Tuning therefore requires a vehicle equipped with an inertial reference, whereas the transfer target does not, because no quantity is re-tuned on it.
The first steps of each recording are excluded from $\mathcal{K}_m$, so that the initialization transient cannot bid the process noise up.

The free parameters are the base-ten logarithms of the standard deviations, grouped by channel, so the swarm searches decades instead of noise values, and the search box spans two decades on either side of the initial guess.
The optimizer is a PSO with twelve particles over forty iterations, which is robust against the flat and ridged objective that the weakly identified noise directions produce.
Tuning integrates with the same explicit Euler scheme at $n=100$ sub-steps that is later deployed, so the tuned noise matches the discrete-time model it is used with.
Wheels reuses the Rigid process noise with the motor-speed measurement disabled, so the two differ in that channel alone.
Rigid, Elastic, and GRU are therefore each evaluated at their own tuned optimum, so the rigid baseline runs at the best process noise the same search could find for it.
The resulting values are reported in Section~\ref{subsec:results_tuning}.

The tuning is repeated three times per variant with independently initialized swarms.
The three runs reach a comparable training objective.
The tuned noise is static, adapting neither over time nor with speed.

\section{Experimental Setup}
\label{sec:setup}

\subsection{Vehicles and Data}
\label{subsec:vehicles}
Both test vehicles are series-production battery-electric vehicles, driven by a single rear-mounted motor through a fixed single-speed transmission.
All estimation results rest on data recorded in test drives with these two vehicles.
Their make and model are withheld for reasons of industrial confidentiality.
\VehA{}, a mid-size sedan, is the development vehicle, whose recordings serve the parameter identification, the tuning of Section~\ref{subsec:tuning}, and the comparison of the four variants.
\VehB{}, a compact SUV, is the transfer target and contributes no data to any identification or tuning step.
Table~\ref{tab:vehicleparams} lists the model parameters of both vehicles, grouped by where their values come from.

The estimator reads its inputs from the controller area network (CAN) bus of the vehicle.
The production signals comprise the four wheel speeds, the yaw rate and the body accelerations of the stability-control unit, the motor speed at its native resolution, the motor torque, the brake pressure, and the steering-wheel angle.
The driven wheels additionally carried externally mounted reference wheel-speed sensors, whose signal served as the training target of the virtual sensor~\cite{schaefke2026virtual}.
The inertial reference is an Automotive Dynamic Motion Analyzer (ADMA-G-ECO+), a GNSS-aided inertial measurement unit sampled at \qty{50}{\hertz}.
The unit is set up with its measured mounting offsets, so it delivers the motion state at the center of gravity.
It receives real-time kinematic (RTK) correction data over a mobile network and is specified with a velocity accuracy of \qty{0.03}{\kilo\meter\per\hour} RMS. In the standstill phases of the test recordings it reads a residual velocity of \qty{0.005}{\meter\per\second} RMS, well below the differences reported in Section~\ref{sec:results}.
Its longitudinal velocity is the ground truth for every error reported in this work.
It never enters the filter, and it is used only to score the estimates and, on the training recordings, to tune the process noise.

The bus channels arrive asynchronously, so each recording is resampled onto a common \qty{50}{\hertz} grid, the filter sampling rate $T_{\mathrm{s}}$ of Section~\ref{subsec:filter}.
When the vehicle is braked to standstill, the sprung mass nods forward and springs back, and the ADMA registers this body motion as a brief negative longitudinal velocity, a genuine motion of the sprung mass but not the travel velocity that longitudinal control acts on.
A first-order Butterworth low-pass at \qty{1.2}{\hertz}, applied forward and backward so that it introduces no phase shift, removes this component and leaves the braking ramp itself in the passband.
All four variants are scored against the same corrected ground truth, and their ranking is the same with and without the \qty{1.2}{\hertz} low-pass.

Deployment on \VehB{} is zero-shot, and the parameters fall into three groups.
The process noise, the measurement noise, the drivetrain trio $I_{\mathrm{EM}}$, $c_{\mathrm{S}}$, $d_{\mathrm{S}}$, and the weights of the virtual sensor are carried over unchanged.
The driveline parameters are confidential and are given as orders of magnitude, \qty{e4}{\newton\meter\per\radian} for the shaft stiffness, \qty{e2}{\newton\meter\second\per\radian} for its damping, and \qty{1}{\kilogram\meter\squared} for the motor inertia.
The geometry, the mass, the dynamic rolling radius, and the gear ratio are vehicle-specific and are taken from the vehicle data sheet or read from the CAN bus of the vehicle.
The CG-to-axle distance is not part of the data sheet of either vehicle and is assumed for both from a near-even axle-load split.
The tire stiffness and the rolling-resistance coefficients were identified from \VehA{} data and fall back to generic default values on \VehB{}, and the remaining Pacejka coefficients are generic on both vehicles.
The yaw inertia is not listed in the data sheet either and is set to a plausible value for each vehicle.
Varying it between \qtyrange{2000}{5000}{\kilogram\meter\squared} changes the RMSE($v_x$) of every variant by less than \qty{0.003}{\meter\per\second}.
No parameter is optimized on \VehB{} data and the only \VehB{} information that enters is the choice among three swarm seeds of comparable training objective.
The transfer target is thus parameterized less precisely than the source vehicle, which makes the comparison of Section~\ref{subsec:results_vehb} a conservative one.

\subsection{Maneuvers and Evaluation Basis}
\label{subsec:maneuvers}
Both campaigns follow the same catalog, organized in three maneuver classes.
Low-speed maneuvers cover creeping, repeated starting and stopping, and parking, the maneuvers of the target application.
Cornering at medium speed, driven as steady circles, checks that the added drivetrain information does not degrade the estimate away from standstill.
Curb crossings, driven forward and backward, load and release the driveline within a few tenths of a second and therefore excite the torsion that separates Elastic from the rigid variants.
To our knowledge, curb crossings have not previously been used to evaluate low-velocity state estimation, although parking functions must operate across them.

The \VehA{} campaign is split into a training set and five held-out test recordings.
The PSO of Section~\ref{subsec:tuning} runs on eight training recordings covering creeping, starting and stopping, circular driving, a parking-garage drive, curb crossings, and one run at \qty{70}{\kilo\meter\per\hour}.
The \VehB{} campaign serves evaluation only.

The evaluation is restricted to the three maneuver classes above, which were fixed before the results were computed and follow the scope of the target application: maneuvering at walking pace, where the production wheel-speed signal degrades.
On \VehA{} this covers all five test recordings, two of them low speed, one cornering, and two curb crossings.
On \VehB{} it covers 18 recordings, 11 of them low speed, three cornering, and four curb crossings.
Since no \VehB{} recording serves identification or tuning, every recording of the three classes is available for evaluation.

\subsection{Metric and Experiments}
\label{subsec:metric}
The reported quantity is the root-mean-square error (RMSE) of the longitudinal velocity against the ADMA reference.
The first \qty{2}{\second} of every recording are discarded, so the initialization transient of the filter does not enter the score.
Each recording belongs to one maneuver class as a whole, and errors are pooled sample by sample within a class: the squared errors of its recordings are concatenated before the root is taken, so every sample carries the same weight regardless of the length of its recording.
Two aggregations are reported for each group, one over the complete recordings and one restricted to the samples where the reference speed stays below \qty{1}{\kilo\meter\per\hour}, that is, \qty{0.278}{\meter\per\second}.
The second aggregation is the target regime of this work, the speed range below \qty{1}{\kilo\meter\per\hour} in which the production wheel-speed sensors lose resolution.
The individual wheel speeds carry no ground truth of their own, so the body velocity $v_x$ remains the evaluated quantity, also for the variant that replaces two wheel-speed measurements.
All reductions are stated relative to Rigid.

\begin{table}[t]
\caption{Model Parameters and Their Sources}
\label{tab:vehicleparams}
\centering
\footnotesize
\setlength{\tabcolsep}{3pt}
\renewcommand{\arraystretch}{1.1}
\begin{tabular}{@{}ll rr@{}}
\toprule
Parameter & Unit & \VehA{} & \VehB{} \\
\midrule
\multicolumn{4}{@{}l}{\textit{From the vehicle data sheet}}\\
$m$              & \unit{\kilogram}                    & \num{2200}    & \num{2150}   \\
$l$              & \unit{\meter}                       & \num{3.0}     & \num{2.8}    \\
$w_{\mathrm{f}}$ & \unit{\meter}                       & \num{1.59}    & \num{1.59}   \\
$w_{\mathrm{r}}$ & \unit{\meter}                       & \num{1.56}    & \num{1.56}   \\
$c_{\mathrm{w}}$ & ---                                 & \num{0.25}    & \num{0.30}   \\
\midrule
\multicolumn{4}{@{}l}{\textit{Measured or assumed in this work}}\\
$I_z$            & \unit{\kilogram\meter\squared}      & \num{4757}    & \num{3900}   \\
$l_{\mathrm{f}}$ & \unit{\meter}                       & \num{1.49}    & \num{1.42}   \\
$A$              & \unit{\meter\squared}               & \num{2.46}    & \num{2.56}   \\
$I_{\mathrm{w}}$ & \unit{\kilogram\meter\squared}      & \num{1.2}     & \num{1.2}    \\
\bottomrule
\end{tabular}

\vspace{2pt}
{\footnotesize\raggedright
The identified tire, rolling-resistance, and driveline parameters are confidential and are therefore not listed. Some parameters are rounded.\par}
\end{table}

\section{Results}
\label{sec:results}
The solver selected by the discretization benchmark E1 in Section~\ref{sec:discretization} is used for every result below.
This section reports the tuned filter parameters (E2) that all following results rest on, the four-way comparison on \VehA{} (E3), and the zero-shot transfer to \VehB{} (E4).
All values are RMSE($v_x$) against the ADMA reference, evaluated on the three maneuver classes of Section~\ref{subsec:maneuvers}.

\subsection{Tuned Filter Parameters (E2)}
\label{subsec:results_tuning}
Table~\ref{tab:pso} reports, for each variant, the process noise returned by the PSO and the measurement noise fixed a priori.
These values are used unchanged in E3 and E4.

The three tuned settings differ by several decades on individual channels rather than by a common scale factor.
The rigid variants place their largest process noise on the motor-speed state, at \qty{11.2}{\radian\per\second}, one to two orders of magnitude above the value that Elastic and GRU receive on the same channel.
Elastic instead keeps the motor speed tight and assigns the torsion angle \qty{3.2e-7}{\radian}, the smallest entry in the table.
GRU carries its largest process noise on the wheel-speed states, the channels that the virtual sensor feeds.

\subsection{Four-Way Comparison on \VehA{} (E3)}
\label{subsec:results_veha}
\begin{figure*}[t]
\centering
\includegraphics[width=\textwidth]{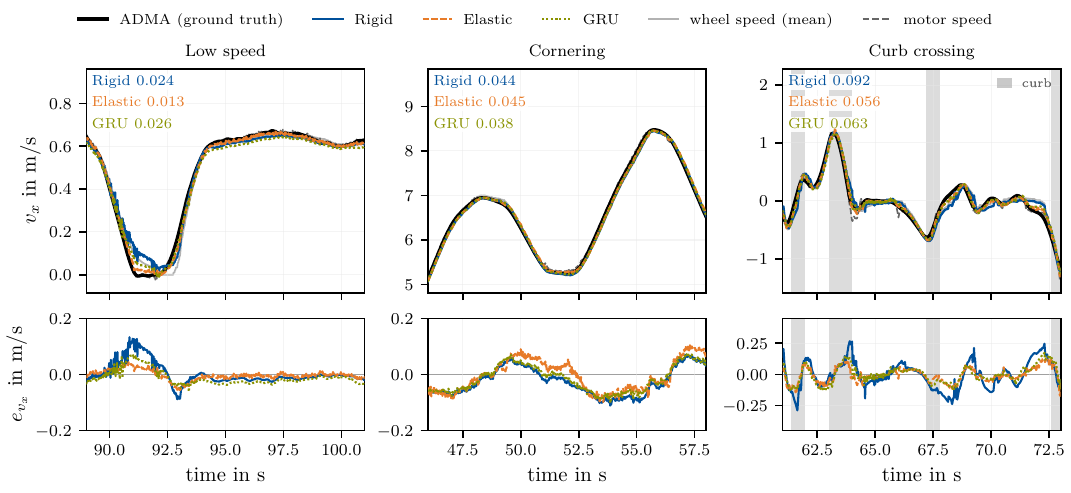}
\caption{Longitudinal velocity $v_x$ against the ADMA ground truth on \VehA{} (E3), shown for three representative scenes (low speed, cornering, curb crossing), each zoomed to a short window.
Top: $v_x$ for Rigid (the rigid-driveline baseline), Elastic (the drivetrain model), and GRU (the virtual sensor), with the velocities derived from the raw wheel speeds and the raw motor speed drawn as thin lines.
Bottom: the signed error $e_{v_x}=\hat{v}_x-v_x$ on per-class axes shared with Fig.~\ref{fig:results_vehb}, \qty{\pm 0.2}{\meter\per\second} for low speed and cornering and \qty{\pm 0.45}{\meter\per\second} for the curb crossings, wide enough to contain every excursion.
In the curb column, the shaded bands mark the curb contacts, located from the ADMA pitch rate.
The box in each upper panel gives the pooled RMSE($v_x$) in \unit{\meter\per\second} over the full recording, not over the window shown.
Wheels is omitted here because it coincides with Rigid; Fig.~\ref{fig:rmse_veha} carries all four variants.}
\label{fig:results_veha}
\end{figure*}

\begin{figure}[t]
\centering
\includegraphics[width=\columnwidth]{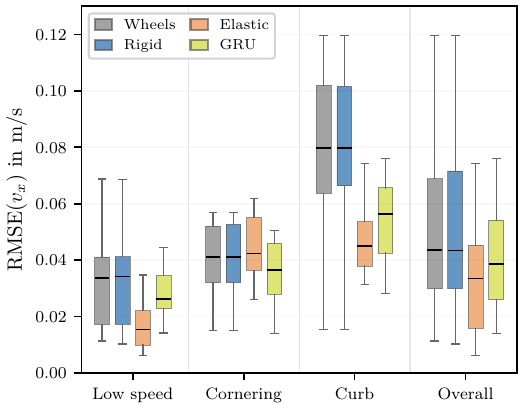}
\caption{Distribution of RMSE($v_x$) per variant on \VehA{} (E3), grouped by maneuver class and overall.
Each box aggregates the per-window RMSE (\qty{10}{\second} windows) over the test recordings of that class. The box spans the interquartile range with the median as a line, the whiskers extend to \num{1.5} interquartile ranges, and points beyond them are not drawn.
The Wheels and Rigid distributions nearly coincide in every class, whereas the Elastic (the drivetrain model) and GRU (the virtual sensor) boxes sit lower in the low-speed and curb classes.}
\label{fig:rmse_veha}
\end{figure}

E3 quantifies the gain of the two proposed estimators on the source vehicle.
Fig.~\ref{fig:results_veha} shows $v_x$ and its error over time for one representative scene per maneuver class, and Fig.~\ref{fig:rmse_veha} the distribution of the windowed RMSE over the test set.
In the sub-\qty{1}{\kilo\meter\per\hour} target regime, Elastic reduces RMSE($v_x$) by \qty{55}{\percent}, from \qty{0.098}{\meter\per\second} for Rigid to \qty{0.044}{\meter\per\second}, and GRU reduces it by \qty{46}{\percent} to \qty{0.052}{\meter\per\second}.
Pooled over all three maneuver classes, the reductions are \qty{34}{\percent} for Elastic, from \qty{0.059}{\meter\per\second} to \qty{0.039}{\meter\per\second}, and \qty{25}{\percent} for GRU, to \qty{0.045}{\meter\per\second}.

Wheels and Rigid agree to within \qty{1}{\percent} in both regimes, with \qty{0.059}{\meter\per\second} for both when pooled and \qty{0.097}{\meter\per\second} against \qty{0.098}{\meter\per\second} below \qty{1}{\kilo\meter\per\hour}.
Adding the motor-speed measurement to the rigid process model thus leaves the accuracy unchanged, so the reductions reported above require the torsion model or the virtual sensor in addition to the motor-speed measurement, not the measurement alone.

The curb class separates the variants most clearly.
Rigid reaches \qty{0.085}{\meter\per\second} there, its weakest class, while Elastic reaches \qty{0.051}{\meter\per\second}, a reduction of \qty{40}{\percent}, and GRU \qty{0.059}{\meter\per\second}.

The cornering class covers the medium-speed portion of the test set.
There GRU reaches the lowest pooled RMSE($v_x$) of the four variants at \qty{0.038}{\meter\per\second}, ahead of Wheels at \qty{0.043}{\meter\per\second}, Rigid at \qty{0.044}{\meter\per\second}, and Elastic at \qty{0.046}{\meter\per\second}.
Elastic is \qty{4}{\percent} above Rigid in this class, on a single recording, while GRU is the most accurate of the four. Neither route therefore trades the low-speed gain for a loss at medium speed.

\subsection{Zero-Shot Transfer to \VehB{} (E4)}
\label{subsec:results_vehb}
\begin{figure*}[t]
\centering
\includegraphics[width=\textwidth]{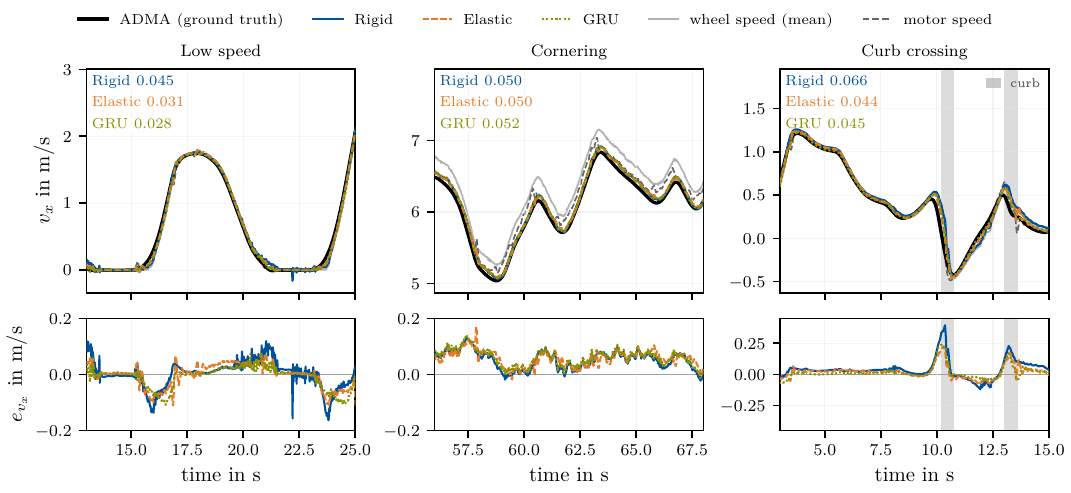}
\caption{Zero-shot cross-vehicle transfer to \VehB{} (E4): the \VehA{} parameters are carried over without re-tuning.
Same layout and variants as Fig.~\ref{fig:results_veha}, shown for three scenes on \VehB{}.}
\label{fig:results_vehb}
\end{figure*}

\begin{figure}[t]
\centering
\includegraphics[width=\columnwidth]{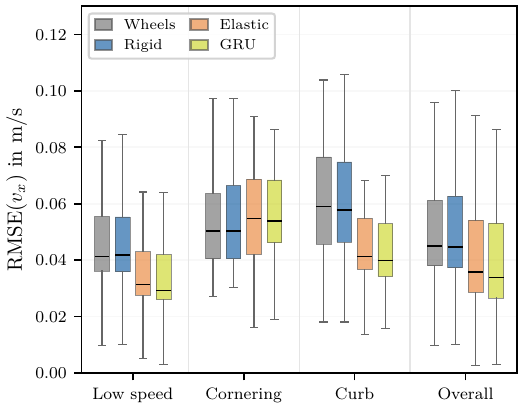}
\caption{RMSE($v_x$) per variant on \VehB{} (E4, zero-shot transfer: \VehA{} parameters, no re-tuning), grouped as in Fig.~\ref{fig:rmse_veha} and on the same axis for comparison. Boxes, medians, and whiskers follow Fig.~\ref{fig:rmse_veha}.}
\label{fig:rmse_vehb}
\end{figure}

The \VehA{} parameters and the trained network are carried over to \VehB{} without re-tuning.
Fig.~\ref{fig:results_vehb} shows $v_x$ and its error over time for three \VehB{} scenes, and Fig.~\ref{fig:rmse_vehb} the distribution of the windowed RMSE.
In the sub-\qty{1}{\kilo\meter\per\hour} regime, GRU reduces RMSE($v_x$) by \qty{52}{\percent}, from \qty{0.066}{\meter\per\second} for Rigid to \qty{0.032}{\meter\per\second}, and Elastic by \qty{42}{\percent} to \qty{0.038}{\meter\per\second}.
Pooled over the three maneuver classes, Elastic reduces the error by \qty{19}{\percent}, from \qty{0.059}{\meter\per\second} to \qty{0.047}{\meter\per\second}, and GRU by \qty{18}{\percent} to \qty{0.048}{\meter\per\second}.
Wheels again stays within \qty{1}{\percent} of Rigid in both regimes.

In the target regime GRU leads in every maneuver class, at \qty{0.029}{\meter\per\second} for low speed, \qty{0.021}{\meter\per\second} for cornering, and \qty{0.042}{\meter\per\second} for the curb crossings, against \num{0.064}, \num{0.046}, and \qty{0.075}{\meter\per\second} for Rigid.
Elastic is second in all three.

The ranking of the two proposed variants is thus reversed relative to \VehA{}: GRU is the stronger of the two in the target regime on the transfer vehicle, Elastic on the source vehicle.
Both remain ahead of the baseline in both regimes and on both vehicles.
Section~\ref{sec:discussion} returns to this.

\begin{table}[t]
\caption{PSO-Tuned Process Noise and Fixed Measurement Noise per Variant (E2)}
\label{tab:pso}
\centering
\footnotesize
\setlength{\tabcolsep}{4pt}
\renewcommand{\arraystretch}{1.15}
\begin{tabular}{@{}ll S[table-format=1.2e-1] S[table-format=1.2e-1] S[table-format=1.2e-1]@{}}
\toprule
State & Unit & {Wheels/Rigid} & {Elastic} & {GRU} \\
\midrule
\multicolumn{5}{@{}l}{\textit{Process noise} $\sigma_{\mm{Q}}$ (PSO-tuned)}\\
$\dot\psi$                 & \unit{\radian\per\second}          & 2.5e-4 & 1.3e0  & 6.9e-3 \\
$v_x$                      & \unit{\meter\per\second}           & 1.3e-1 & 1.4e-2 & 4.2e-2 \\
$v_y$                      & \unit{\meter\per\second}           & 3.4e-3 & 1.2e-2 & 1.6e-4 \\
$\omega_{\mathrm{fl\dots rr}}$ & \unit{\radian\per\second}     & 5.6e-2 & 5.9e-2 & 3.6e-1 \\
$\omega_{\mathrm{EM}}$     & \unit{\radian\per\second}          & 1.12e1 & 1.1e-1 & 2.0e-1 \\
$\Delta\theta$             & \unit{\radian}                     & {\tno} & 3.2e-7 & {\tno} \\
\midrule
\multicolumn{5}{@{}l}{\textit{Measurement noise} $\sigma_{\mm{R}}$ (fixed a priori, all variants)}\\
$\dot\psi$                 & \unit{\radian\per\second}          & \multicolumn{3}{c}{\num{0.01}} \\
$a_x,\,a_y$                & \unit{\meter\per\second\squared}   & \multicolumn{3}{c}{\num{0.2}}  \\
$\omega_{\mathrm{fl\dots rr}}$ & \unit{\radian\per\second}     & \multicolumn{3}{c}{\num{0.1}$^{\ddagger}$}  \\
$\omega_{\mathrm{EM}}$     & \unit{\radian\per\second}          & \multicolumn{3}{c}{\num{0.1}$^{\dagger}$} \\
\bottomrule
\end{tabular}

\vspace{2pt}
{\footnotesize\raggedright
Diagonal standard deviations of $\mm{Q}$ and $\mm{R}$. The process noise is added once per filter step of $T_{\mathrm{s}}=\qty{20}{\milli\second}$, so $\sigma_{\mm{Q}}$ carries the unit of its state; \tno~marks the torsion angle, which is inert outside Elastic (Section~\ref{subsec:variants}).\\
$^{\dagger}$~Disabled in Wheels ($R\!\to\!\infty$).\quad $^{\ddagger}$~Fed by the virtual sensor in GRU.\par}
\end{table}

\section{Discussion}
\label{sec:discussion}
Elastic and GRU attack the same deficit, the degraded wheel-speed signal near standstill, along different routes~(Figs.~\ref{fig:rmse_veha} and~\ref{fig:rmse_vehb}).
A rigid driveline slaves the motor speed to the wheels through a kinematic identity that the real drivetrain violates, and the rigid model can absorb that mismatch only by loosening the motor-speed state itself.
The tuned process noise of Wheels and Rigid does exactly this, so the motor-speed measurement no longer constrains the wheel states, which is why the two agree to within \qty{1}{\percent}.
The explicit drivetrain state removes that identity, making the motor speed a torque-driven state of its own whose high-resolution signal informs $v_x$ exactly where the production wheel-speed sensors degrade.
The effect is largest where the wheels are disturbed most, on the curb class of \VehA{}, where Elastic reaches \qty{0.051}{\meter\per\second} against \qty{0.085}{\meter\per\second} for Rigid. Over all classes in the sub-\qty{1}{\kilo\meter\per\hour} target regime, Elastic lowers the error by \qty{55}{\percent}.
The virtual sensor attacks the same degradation one level earlier, on the signal itself.
It maps the production wheel speeds and the motor speed to corrected wheel velocities without representing the drivetrain at all. On \VehB{} it leads in every maneuver class below \qty{1}{\kilo\meter\per\hour} and lowers the pooled error in this target regime by \qty{52}{\percent}.
Elastic, in contrast, carries the driveline parameters of the source vehicle to a target vehicle whose gear ratio differs by a factor of \num{1.6}.
These parameters are laborious to identify, so a dedicated campaign per target vehicle is exactly what the transfer is meant to avoid.
The virtual sensor instead maps signals that carry no such parameters.
This offers a plausible explanation for the reversed ranking of Section~\ref{subsec:results_vehb}, although the present data do not isolate it.
In the cornering class (Fig.~\ref{fig:rmse_veha}), which covers the medium-speed portion of the \VehA{} test set, GRU is the most accurate of the four variants and Elastic stays within \qty{0.002}{\meter\per\second} of the baseline.
Neither route therefore trades the low-speed gain against accuracy elsewhere.

Both proposals reach a similar accuracy in the target regime, so the choice between them rests on practical differences.
Elastic requires the driveline parameters of the target vehicle, transferred here instead of measured, while GRU requires recordings from a comparable platform to train on.
Neither is tuned or trained on target-vehicle data.
Elastic estimates a physical quantity, the shaft twist, which can be inspected and bounded, whereas GRU returns a point estimate whose measurement noise must be fixed a priori, so the filter cannot distinguish a confident output from an extrapolated one.
The computational cost is identical as deployed, because all four variants integrate at the same $n=100$ sub-steps.
The process-model integration that dominates each sigma-point propagation runs at a real-time factor of \numrange{750}{800} (Section~\ref{sec:discretization}), the variants differ by at most one additional state, and the network evaluation is small by comparison.
A practical GRU deployment would not be bound to that count, since the rigid process model it runs on is stable from $n=40$ (Appendix~\ref{app:euler_sweep}), whereas Elastic requires $n=100$ for stability.
Elastic is therefore the preferable route when the driveline parameters of the target vehicle are accessible and the estimate must remain interpretable. GRU is preferable when those parameters are uncertain and recordings from a comparable platform are available.

The proposed model rests on the drivetrain parameters.
Because the target maneuvers excite the driveline too little to resolve it, the shaft stiffness is only weakly identifiable in this data. A fit of the drivetrain model alone runs into the upper end of the physically plausible range for a half-shaft, the error of the complete filter falls monotonically with increasing stiffness within that range without reaching an optimum, and the measured torsion rate shows no resonance peak. None of the three argues against modeling the compliance, only against identifying its stiffness.
The stiffness is therefore not a fitted optimum but set by matching the measured motor-to-wheel speed deviation on load changes, and the shaft damping and the motor inertia carry the same provenance.
Elastic stays ahead of Rigid across the range of stiffnesses the data support, so the gain follows from the explicit drivetrain state, not from the exact parameter values, and the orders of magnitude of Section~\ref{subsec:vehicles} are enough to reproduce it.
Adding gear backlash to the shaft model brought no improvement and was not pursued further.
The yaw inertia is likewise set to a plausible value rather than identified, but it is common to all four variants and moves the RMSE($v_x$) of every variant by less than \qty{0.003}{\meter\per\second} over the range of Section~\ref{subsec:vehicles}.
Resolving the two driven wheels exactly instead of the lumped form of~\eqref{eq:inertia_rigid} changes the RMSE($v_x$) of every variant by less than \qty{0.002}{\meter\per\second} and leaves the ranking unchanged.

Two limitations bound these results.
First, the tuned noise is static, which leaves a speed-scheduled or online-adapted process noise as the natural next step.
Second, the virtual sensor enters the filter as a point estimate, so its measurement noise is fixed a priori, and the same network feeds both rear-wheel channels, whose correlated errors the diagonal $\mm{R}$ ignores.
Deriving the noise from the validation error of the network, or from a network that predicts its own uncertainty, would let the filter weight the virtual measurement by its reliability.

\section{Conclusion}
\label{sec:conclusion}
Conventional velocity estimation relies on production wheel-speed sensors, which lose resolution below \qty{1}{\kilo\meter\per\hour}, so the estimate degrades although driver assistance depends on it.
We integrate the high-resolution motor speed of an electric drivetrain into a low-velocity double-track UKF along two routes, an explicit torsional drivetrain model (Elastic) and a data-driven virtual wheel-speed sensor (GRU). Both are compared against a rigid-driveline baseline (Rigid) and a wheel-speed-only configuration (Wheels).
In real-world test drives on \VehA{}, both estimators lower RMSE($v_x$) in the target regime below \qty{1}{\kilo\meter\per\hour} relative to Rigid, by \qty{55}{\percent} for Elastic and \qty{46}{\percent} for GRU.
Carried over to \VehB{} without re-tuning, Elastic lowers the error by \qty{42}{\percent} and GRU by \qty{52}{\percent}.
Wheels stays within \qty{1}{\percent} of Rigid on both vehicles, so the additional sensor alone leaves the accuracy unchanged and the gain requires the torsion model or the virtual sensor.
The zero-shot transfer carries over the noise covariances, the drivetrain parameters, and the trained network unchanged, and no quantity is fitted on data of the target vehicle.
The evaluation includes curb crossings, a maneuver class that parking functions must handle and that, to our knowledge, previous velocity-estimation studies have not covered.

The low-velocity model, discretized by a scheme chosen in a systematic comparison of explicit and semi-implicit integrators, stays stable down to standstill at real-time cost.
Integrating the motor speed reduces the error in the sub-\qty{1}{\kilo\meter\per\hour} regime without degrading accuracy at medium speed. In the cornering class of the \VehA{} test set, Elastic stays within \qty{0.002}{\meter\per\second} of Rigid and GRU is more accurate than Rigid.
Both estimators transfer to a second vehicle without re-tuning and keep their advantage in the target regime.
Elastic therefore suits a target vehicle with accessible driveline parameters and an interpretability requirement, while GRU suits one whose parameters are uncertain but for which recordings from a comparable platform exist.
The gain in this regime directly serves the functions that operate there, such as automated parking and anti-jerk drivetrain control.
Promising directions for future research are process noise that adapts with speed and virtual sensors that report their own uncertainty so that the filter can weight them.

\appendices

\section{Closed-Loop Step-Size Sweep}
\label{app:euler_sweep}
The discretization benchmark of Section~\ref{sec:discretization} selects the integration scheme open loop, on a single synthetic test maneuver.
This appendix reports its closed-loop counterpart. A sweep of the sub-step count inside the running filter locates the stability limit of each driveline form and shows how much accuracy finer stepping buys.
For the elastic-driveline UKF (Elastic) and the rigid-driveline baseline (Rigid), the adopted Euler-100 process noise is deployed at Euler sub-step counts from 20 to 160 on the \VehA{} test set, so that the discretization effect is isolated.
Each sub-step count is scored by RMSE($v_x$) over 15 runs (three tuned seeds times five test recordings); a run counts as diverged when its RMSE is non-finite or exceeds \qty{0.2}{\meter\per\second}.

\begin{figure}[t]
\centering
\includegraphics{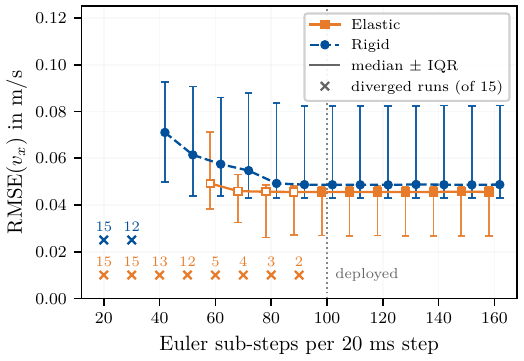}
\caption{Closed-loop Euler step-size sweep on the \VehA{} test set: median RMSE($v_x$) and interquartile range of the converged runs against the number of Euler sub-steps per \qty{20}{\milli\second} filter step, for Elastic and Rigid at the adopted Euler-100 process noise. Open markers denote sub-step counts with diverged runs, and the crosses count them. The dotted line marks the deployed operating point $n=100$.}
\label{fig:euler_sweep}
\end{figure}

Fig.~\ref{fig:euler_sweep} summarizes the sweep.
The rigid-driveline baseline is stable from $n=40$ sub-steps, whereas the elastic driveline requires $n=100$ for every run to converge.
The explicit torsion state therefore costs about \num{2.5} times the sub-steps of the rigid model.
The accuracy is flat above these limits, with the median RMSE($v_x$) holding near \qty{0.046}{\meter\per\second} for Elastic and \qty{0.049}{\meter\per\second} for Rigid from $n=100$ to $n=160$.
The deployed $n=100$ therefore sits directly at the stability limit and gains nothing from finer stepping.
The diverging runs occur only on the curb-crossing maneuvers, whose load changes excite the drivetrain torsion, consistent with the stiff dynamics of Sections~\ref{sec:model} and~\ref{sec:discretization}.

This stability floor is a property of the stiff dynamics, not of the tuning.
Re-tuning the elastic process noise directly at the coarse step does not move it.
Of three seeds tuned at Euler-60, two diverge, and the stable seed is nearly twice as poor in the target regime (\qty{0.087}{\meter\per\second} against \qty{0.044}{\meter\per\second} for the deployed configuration).
Of three seeds tuned at Euler-80, one diverges (pooled RMSE \qty{0.33}{\meter\per\second}).
Deployed at Euler-100, all six re-tuned settings are stable.
The sweep thus confirms the Euler-100 deployment anchor of Section~\ref{sec:discretization} in closed loop and quantifies the cost of the explicit torsion model.

\section*{Acknowledgment}
The two vehicle illustrations in Fig.~\ref{fig:concept}(c) were generated with Google Gemini from text prompts and are illustrative only.
The authors used Anthropic's Claude in the preparation of this article.
All content was reviewed and verified by the authors.

\bibliographystyle{IEEEtran}
\bibliography{literature}

\end{document}